\UseRawInputEncoding
\documentclass[aps,prx,twocolumn,superscriptaddress,longbibliography,nofootinbib,eprint,floatfix]{revtex4-2}
\usepackage[english]{babel}
\usepackage{listings}
\usepackage{tikz}
\usepackage{subcaption}
\usepackage{float}
\usepackage{comment}
\usepackage{minted}
\usepackage{xcolor}
\usepackage{fvextra}
\usepackage{amsmath}
\usepackage[normalem]{ulem}
\usepackage{amssymb}
\usepackage{graphicx}
\usepackage{physics}
\usepackage{braket}
\usepackage{algorithm}
\usepackage{algpseudocode}
\usepackage{tcolorbox} 
\usepackage[colorlinks=true, allcolors=blue]{hyperref}
\usepackage[letterpaper,top=2cm,bottom=2cm,left=3cm,right=3cm,marginparwidth=1.75cm]{geometry}

\definecolor{bg}{rgb}{0.12, 0.12, 0.12}
\definecolor{LightGray}{gray}{0.9}
\definecolor{darkgreen}{rgb}{0, 0.5, 0}
\definecolor{violet}{RGB}{107, 29, 129}

\newcommand{\dist}{\text{dist}}

\usetikzlibrary{positioning, arrows.meta, shapes.misc, shapes.geometric, calc, fit}
\tikzset{block/.style={rectangle, draw, fill=blue!20, minimum height=3em, minimum width=6em, text centered, text width=5.5em}, arrow/.style={-Latex}, line/.style={draw, -latex'}}

\begin{document}



\title{Quantum circuit partition as a maze: emerging percolation\\ transition via path finding}

\author{P. Zentilini}
\affiliation{Department of Physics Aldo Pontremoli, Universit\`a degli Studi di Milano, Italy}
\affiliation{Istituto di Fotonica e Nanotecnologie, Consiglio Nazionale delle Ricerche, Italy}

\author{M. Guatto}
\thanks{These authors contributed equally to this work.}
\affiliation{Forschungszentrum J\"ulich, Institute of Quantum Control (PGI-8), D-52425 J\"ulich, Germany}
\affiliation{Institute for Theoretical Physics, University of Cologne, 50937 Cologne, Germany}

\author{F. Preti}
\thanks{These authors contributed equally to this work.}
\affiliation{J\"ulich Supercomputing Centre, Helmholtz AI, 52428 J\"ulich, Germany}

\author{D. Arya}
\affiliation{Forschungszentrum J\"ulich, Institute of Quantum Control (PGI-8), D-52425 J\"ulich, Germany}
\affiliation{Institute for Theoretical Physics, University of Cologne, 50937 Cologne, Germany}

\author{F. A. C\'ardenas-L\'opez}
\affiliation{Forschungszentrum J\"ulich, Institute of Quantum Control (PGI-8), D-52425 J\"ulich, Germany}

\author{F. Motzoi}
\affiliation{Forschungszentrum J\"ulich, Institute of Quantum Control (PGI-8), D-52425 J\"ulich, Germany}
\affiliation{Institute for Theoretical Physics, University of Cologne, 50937 Cologne, Germany}

\author{E. Prati}
\affiliation{Department of Physics Aldo Pontremoli, Universit\`a degli Studi di Milano, Italy}
\affiliation{Istituto di Fotonica e Nanotecnologie, Consiglio Nazionale delle Ricerche, Italy}

\begin{abstract}
In quantum circuit optimization, circuit partitioning enables the optimization process to be parallelized across multiple devices.
Each device is responsible for either reducing the number of selected gates or simplifying the local circuit structure.
Most existing approaches to circuit partitioning are quantum-distribution-oriented and rely on splitting CNOT gates by introducing mid-circuit measurements and qubit resets.
Currently, there is no criterion to determine how a circuit can be optimally partitioned without removing the CNOT gates for circuit optimization purposes.
To address this challenge, we formalize the partition problem as a cutting path through a maze, where the CNOT gates represent the walls. 
We show that the existence of such a path separates quantum circuits into two classes through a percolation phase transition. 
In particular, it distinguishes a partitionable regime from a nonpartitionable one, arising from qubit permutations.
Such permutations are generated by simulated annealing.
We analyze its effect on the CNOT cluster from the perspective of network science and distribution analysis. 
Our results show that partitioning into two CNOT clusters is possible when the number of CNOTs is almost equal to the number of qubits. 
Based on this observation, we provide a scalable and practical criterion for identifying whether such a partition exists. 
Overall, our framework provides theoretical and numerical insight into circuit partitioning
within quantum circuit optimization, forming the basis for algorithmic development.
\end{abstract}

\keywords{Percolation, phase transition, quantum circuit partitioning}

\maketitle

\section{Introduction}

Quantum algorithms promise to revolutionize present computing capabilities, with applications ranging from quantum chemistry simulations \cite{lanyon2010towards} to machine learning training \cite{Liu2024, preti2026gradients}. 
Despite substantial improvements in the performance of the QPUs, current devices remain far from the error rates required to run complex algorithmic routines \cite{Ruane2025QuantumIndex}. 
Since such routines typically require many qubits and deep circuits, circuit simplification and hardware-aware optimization remain essential for practical quantum advantage \cite{nam2018automated, ruiz2025quantum}.
 
Advanced error-correction codes are substantially improving logical error rates \cite{Acharya2025, guatto2025, Sivak_2023}, while highly efficient compilation and transpilation methods, including measurement-based approaches, are enabling increasingly shorter, high-fidelity circuits \cite{BQSKIT, Ender2023, nemirovsky2025efficientcompilationquantumcircuits, Kissinger_2020, Preti2024hybriddiscrete, wang2025circuitdesignstarshapedspinqubit, moro2021quantum, banfi2025transpiling, kremer2024practical, corli2026measurement, corli2025gauge}. 
There, the relevant optimization goal depends on the stage of quantum technology development considered. 
In fault-tolerant architectures, minimizing non-Clifford resources, most commonly the T-count or T-depth is critical, as non-Clifford gates are typically much more resource-intensive than Clifford gates in more advanced quantum error-correction schemes, such as surface-code-based architectures \cite{Amy_2014, akahoshi2024partially}. 
In contrast, for NISQ devices, reducing the number of two-qubit gates, often reported as the CNOT count after decomposition, is typically a primary goal, as the two-qubit gates are primarily responsible for the noise impact \cite{xu2025optimizing}.
In addition, hardware-aware qubit allocation and alignment are likewise crucial \cite{zulehner2018efficientmethodologymappingquantum, Siraichi2018}.

From this perspective, efficient circuit partitioning is relevant both for present-day NISQ workflows and for the long-term development of large-scale quantum compilation and transpilation pipelines. 
In compilation and transpilation, it serves as a tool to simplify the  resynthesis of large circuits \cite{BQSKIT, wu2020qgo}, whereas in quantum distribution it is used to divide computations into separately executable subcircuits \cite{tejedor2025distributed, tamura2026decompositionmultiqubitgatescircuit, burt2026multilevel}. 
Since cutting entangling gates is acceptable in the latter setting but generally undesirable in the former, the central question is whether a quantum circuit can be partitioned without cutting CNOT gates, and under which criterion this is possible.

To address this, we discuss a framework that formalizes the partitioning of quantum circuits in the context of circuit optimization by a cut as a path in a maze problem, where CNOT gates represent walls. 
By forcing such cutting path to separate the circuit into two subcircuits balanced in both the number of entanglement gates and qubits, a percolation between two classes of quantum circuits arises. 
We then define a criterion based on the characteristics of this path to identify the equivalence class to which a given quantum circuit belongs.

The discovered numerical percolation places the problem in the broader context of critical phenomena in the space of quantum circuits, such as measurement-induced phase transitions \cite{Skinner2019, Li2018quantumzeno, Jian2020, Nahum2021, Lavasani2021}, topological transitions in ZX graphs \cite{Buznach2025}, many-body localization transitions with random time-periodic quantum circuits \cite{sunderhauf2018localization}, the threshold theorem in quantum error correction \cite{aharonov1999faulttolerantquantumcomputationconstant, KITAEV20032}, to name a few. 
Within this perspective, the ability to partition a quantum circuit without cutting any CNOT gates can be viewed as the first phase of a percolation transition.
The criterion defining this stage depends on the ratio between the number of entangling gates and the number of qubits.
For such a partition to be possible, this ratio must scale maximally linearly.

This work is divided in six sections.
In Section \ref{section:background}, we describe the background regarding the link between percolation, network science and quantum circuits.
In Section \ref{section:wire&maze}, we introduce the dataset we deal with and the wire and the maze representations.
Section \ref{section:qubit reordering} analyzes both from a mathematical and numerical point of view the use of simulated annealing to simplify the layout of the quantum circuit before the cut.
In Section \ref{section:phase transition} we show the results of the partitioning heuristic which leads to a percolation behaviour. Section \ref{section:conclusions} summarizes our conclusions.

\section{Background on percolation and critical phenomena in quantum circuits}
\label{section:background}

Recent advances in the study of critical phenomena in quantum circuits have been largely driven by a better understanding of the competition between unitary dynamics and local measurements. 
While unitary gates typically promote entanglement growth, frequent measurements tend to suppress it.
In monitored many-body circuits, low measurement rates can support a volume-law entangled phase, whereas sufficiently frequent measurements drive the system toward an area-law phase~\cite{li2018quantum}.
This transition admits a simple geometric interpretation in terms of the spacetime tensor-network graph associated with the monitored circuit. 
For a fixed circuit geometry and wire ordering, the entanglement can be related, in suitable limits such as the minimal-cut description of the zeroth R\'enyi entropy, to the cost of an optimal cut through this network. 
Measurements effectively remove links from the network, so that the corresponding minimal-cut problem can be mapped to a classical percolation problem~\cite{Skinner2019}.
In Refs.~\cite{choi2020quantum}, random monitored circuits are mapped to effective classical statistical-mechanics models. 
While these models reduce to bond percolation in the large local Hilbert-space dimension limit, finite-dimensional circuits generically exhibit distinct critical behavior and belong to a different universality class \cite{Jian2020,zabalo2020critical}.
Connectivity-driven transitions also arise in measurement-only circuits, topologically monitored dynamics, and ZX-calculus representations of Clifford circuits \cite{jonay2021triunitary,lavasani2021measurement,Buznach2025}. 
These examples already suggest that changes in connectivity alone can induce transitions in entanglement scaling, further highlighting the link between classical percolation theory and quantum circuits.

\textbf{Percolation theory} studies the emergence of large-scale connectivity in disordered systems. 
In site-percolation, each site in the lattice is occupied with probability $p$ and empty with probability $1-p$.
For small $p$, the occupied sites typically form only small, isolated clusters, while as $p$ increases, these clusters grow and merge. 
The system is said to percolate when a cluster crosses the entire system, connecting two opposite sides of the lattice. 

In one dimension, a chain of length $L$ forms only if all $L$ sites are occupied. Therefore the percolation probability is given by
\begin{equation}
    \Pi(p,L)=p^L .
\end{equation}
In the thermodynamic limit, $\Pi(p,L)\to 0$ for all $p<1$, while $\Pi(1,L)=1$. 
Thus, the one-dimensional threshold is $p_c=1$, so no non-trivial transition occurs.
However, in two dimensions, alternative paths make a transition possible. 
Indeed, near the critical point, the percolation probability of a finite system of linear dimension $L$ obeys
\begin{equation}
    \Pi(p,L)=\Phi\!\left[(p-p_c)L^{1/\nu}\right],
\end{equation}
where $\Phi$ is a scaling function and $\nu=4/3$ for two-dimensional percolation. 
The threshold $p_c$ depends on the lattice geometry and connectivity. 
For nearest-neighbor percolation, $p_c \sim 0.5927$ on the square lattice and $p_c \sim 0.6973$ on the honeycomb lattice \cite{djordjevic1982site}.

Percolation also occurs in the context of random graphs. 
In the Erd\"os-R\'enyi model $G(N,p_e)$, for example, edges are present independently with probability $p_e$, yielding an average degree
\begin{equation}
    \langle k\rangle=(N-1)p_e\simeq Np_e .
\end{equation}
In such a model, a phase transition occurs when a giant connected component begins to emerge. This happens at the critical threshold $p_e \simeq \frac{1}{N}$ corresponding to an average node degree $\langle k\rangle=1$ \cite{erdHos1960evolution}.
Several other network science models exhibit such phase transitions, such as the Configuration model \cite{Bollobas1980} or the Barabasi-Albert model \cite{Barabasi1999, barabasi2016network} that describes preferential attachment.

\textbf{Network science} provides a natural framework for describing quantum circuits and quantum many-body systems, as both can be represented in terms of interactions on graphs. 
In this picture, quantum circuits define directed graphs encoding causal structure, allowing the use of graph-based and generative methods for circuit optimization and compilation \cite{beaudoin2024altgraphredesigningquantumcircuits}. 
Beyond this structural analogy, several key concepts from network science have direct quantum counterparts. 
For instance, classical consensus and opinion-dynamics models extend to quantum networks through gossip and symmetrization protocols, where local interactions drive the emergence of global agreement \cite{Mazzarella_2015,Ticozzi_2016}. 
Similarly, centrality measures can be generalized using quantum information concepts, with network entanglement quantifying the role of nodes in mediating correlations and information flow \cite{huang2024network_entanglement}. 
. 
By leveraging these concepts from network science, we provide a characterization of the percolation that emerges in our study.

\section{Description of the dataset and representations of quantum circuits}
\label{section:wire&maze}

In this Section, we describe the main characteristics of the quantum circuits discussed.
We then introduce the maze representation and the wire representation, which are essential for transforming the circuit partitioning problem into a maze-traversal problem, also clarifying the type of cutting-paths we intend to make.

\subsection{Construction of the synthetic dataset of quantum circuits}
\label{sec:dataset}

The dataset consists of a set of randomly generated quantum circuits acting on up to $N_q=100$ qubits.
They are created upon a set of discrete gates with a fixed number of entanglement operations.
Each circuit is initialized in the computational ground state $\ket{0}^{\otimes N_q}$ and is built sequentially by appending gates randomly until a predetermined number of two-qubit gates is reached, which is fixed in the range $10$ to $100$, while the total number of single qubit gates can vary.
At each stage of construction, a gate is sampled uniformly and randomly from the set $$G = \{ X, Y, Z, H, CNOT, Rz(\theta), T, T/2 \}$$
The explicit form of each gate can be found in Appendix \ref{appendix:gates}.
The set consists of Clifford operations in combination with fixed-angle rotations that introduce uncontrolled non-Clifford elements.
For single-qubit gates, the target qubit is chosen uniformly and randomly from the $N$-qubit register.
For two-qubit gates, an ordered pair of distinct qubits $(i,j)$ is sampled uniformly and randomly, defining the control and target qubits of the controlled X -- CNOT -- operations.
No connectivity constraints are imposed, corresponding to an all-to-all interaction graph.
Circuit generation proceeds until exactly $N_{c}$ CNOT gates have been applied.
Single-qubit gates do not contribute to the stopping criterion and can be applied an arbitrary number of times between entanglement operations.
As a result, the circuit depth is variable, while the number of entanglement gates is fixed across the board.

\begin{figure*}[t]
    \centering
    \includegraphics[width=\textwidth]{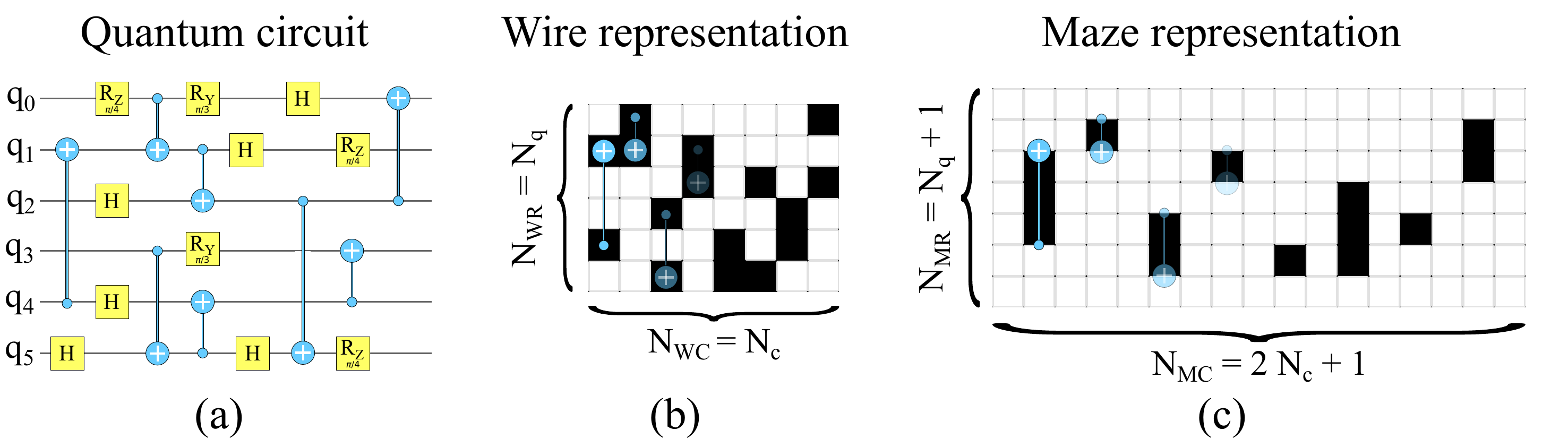}
    \caption{\raggedright Wire representation and maze representation compared to a quantum circuit addressed in this work.
    (a) Example of a randomly generated quantum circuit.
    (b) The corresponding wire representation is constructed by retaining only the CNOT gates. 
    The rows correspond to qubit wires and the columns correspond one-to-one to the CNOT gates of the circuit. 
    A cell is black if the vertical line segment of a CNOT gate intersects that qubit wire, including the wires corresponding to the control and target qubits.
    (c) Maze representation of the circuit. 
    Rows correspond to the spaces between adjacent qubits, augmented by empty padding rows at the top and bottom. 
    Columns alternate between empty columns and columns associated with CNOT gates, with additional empty padding columns at the left and right boundaries, yielding a total of $2\,N_{c}\,+\,1$ columns.
    A black cell indicates that the corresponding CNOT gate spans the given inter-qubit space.}
    \label{fig:maze representation}
\end{figure*}

\subsection{Wire and maze representation of a quantum circuit}
\label{wire&maze}

To formulate the partition of a circuit as a maze-cutting problem, an adequate representation of the circuit is needed.
Quantum circuits allow for many possible representations, each of which highlights different structural properties. 
Common examples include directed acyclic graphs (DAGs), such as those used in Qiskit \cite{javadi2024quantum}, and reduced interaction graphs that maintain only the entangling gates \cite{daei2020optimized}.

In this work, we introduce two geometric interpretations of quantum circuits that reformulate circuit partitioning as a maze-traversal problem, while preserving the spatial and temporal order of operations.
The pseudo-code for generating such representations is provided in Algorithm~\ref{alg:wire-maze-representations}.

\begin{algorithm}[H]
\caption{Construction of $\mathcal{W}$ and $\mathcal{M}$}
\label{alg:wire-maze-representations}
\begin{algorithmic}[1]
\Require ordered CNOT list $\mathcal{C}=\{(c_k,t_k)\}_{k=1}^{N_c}$, where $c_k$ and $t_k$ are the control and target qubits of the $k$-th CNOT
\Ensure $\mathcal{W}\in\{0,1\}^{N_q\times N_c}$ and $\mathcal{M}\in\{0,1\}^{(N_q+1)\times(2N_c+1)}$
\Statex
\State Initialize $\mathcal{W}\gets 0_{N_q\times N_c}$
\State Initialize $\mathcal{M}\gets 0_{(N_q+1)\times(2N_c+1)}$
\For{$k=1$ to $N_c$}
    \State $a \gets \min(c_k,t_k)$
    \State $b \gets \max(c_k,t_k)$
    \For{$i=a, b$}
        \State $\mathcal{W}_{i,k}\gets 1$
    \EndFor
    \State $j \gets 2k$
    \For{$r=a+1$ to $b$}
        \State $\mathcal{M}_{r,j}\gets 1$
    \EndFor
\EndFor
\State \Return $\mathcal{W},\mathcal{M}$
\end{algorithmic}
\end{algorithm}

We first introduce the wire representation $\mathcal{W}$, illustrated in Figure~\ref{fig:maze representation} (b), which provides a direct encoding of the target qubit and control of the CNOT gates of a quantum circuit.
In this representation, the circuit is mapped onto a two-dimensional grid with a number of rows $N_{WR} = N_{q}$ and a number of columns $N_{WC} = N_{c}$.
Each row is in one-to-one correspondence with a wire of qubits in the circuit, preserving the circuit ordering of the CNOT gates.
Each grid element takes the value $1$ if it corresponds to the presence of either a control or a target qubit, and $0$ otherwise.
Each column corresponds to an entanglement gate, and the grid entries encode the positions of the control and target qubits along the wires. 

We then introduce the maze representation $\mathcal{M}$, depicted in Figure~\ref{fig:maze representation} (c), which places the circuit in a two-dimensional grid that can be interpreted as a maze where the CNOT operations constitute the walls.
The grid has a number of rows $N_{\text{MR}} = N_{q} + 1$ and a number of columns $N_{\text{MC}} = 2\,N_{c} + 1$.
The $N_{q}-1$ inner rows correspond to the spaces between adjacent qubit wires, while the remaining two rows provide a blank fill above the top wire and below the bottom wire.
Note that, unlike the $\mathcal{W}$ representation, 
permuting the rows of $\mathcal{M}$ would rearrange the inter-wire spaces without consistently reordering the wires themselves, and the resulting maze would no longer correspond to a valid circuit layout.
Each grid element takes on the value $1$ if it corresponds to a point where a CNOT gate intersects the space between two wires and $0$ otherwise.
Also in this case the spatial ordering of the qubit wires and the chronological ordering of operations are preserved. 
Each entanglement gate occupies a dedicated column, and consecutive columns of entanglement gates are separated by an empty column.
Furthermore, the grid is surrounded by empty padding (including the aforementioned top and bottom padding rows), which provides free space for routing.
This alternating structure of gate columns and empty columns, together with the surrounding padding, guarantees the existence of at least one valid and continuous path connecting the left and right edges of the maze.
Such a construction leads to the following observation. 
Although a path of this type always exists, its mere existence is not sufficient to obtain a non-trivial partition of the circuit, since not all paths generate subcircuits that can be optimized independently. 
Let us therefore introduce a central observation for this work: a sufficient, but not strictly necessary, condition for achieving this goal is that the cutting path be monotonic both in time  -- in the sense that it never moves backward along the time axis -- and along the vertical axis.
\begin{tcolorbox}[
    title=Definition 1, 
    colback=violet!5!white, 
    colframe=violet!75!black,  
    arc=2mm                  
]
A maze/circuit cutting path is \textbf{\textcolor{violet}{monotonic}} on the maze if, for every line (qubit wire) between two maze rows, by drawing a horizontal line on that row, this intersects the path only once.
\end{tcolorbox}
\begin{figure*}[t]
    \centering
    \includegraphics[width=0.95\textwidth]{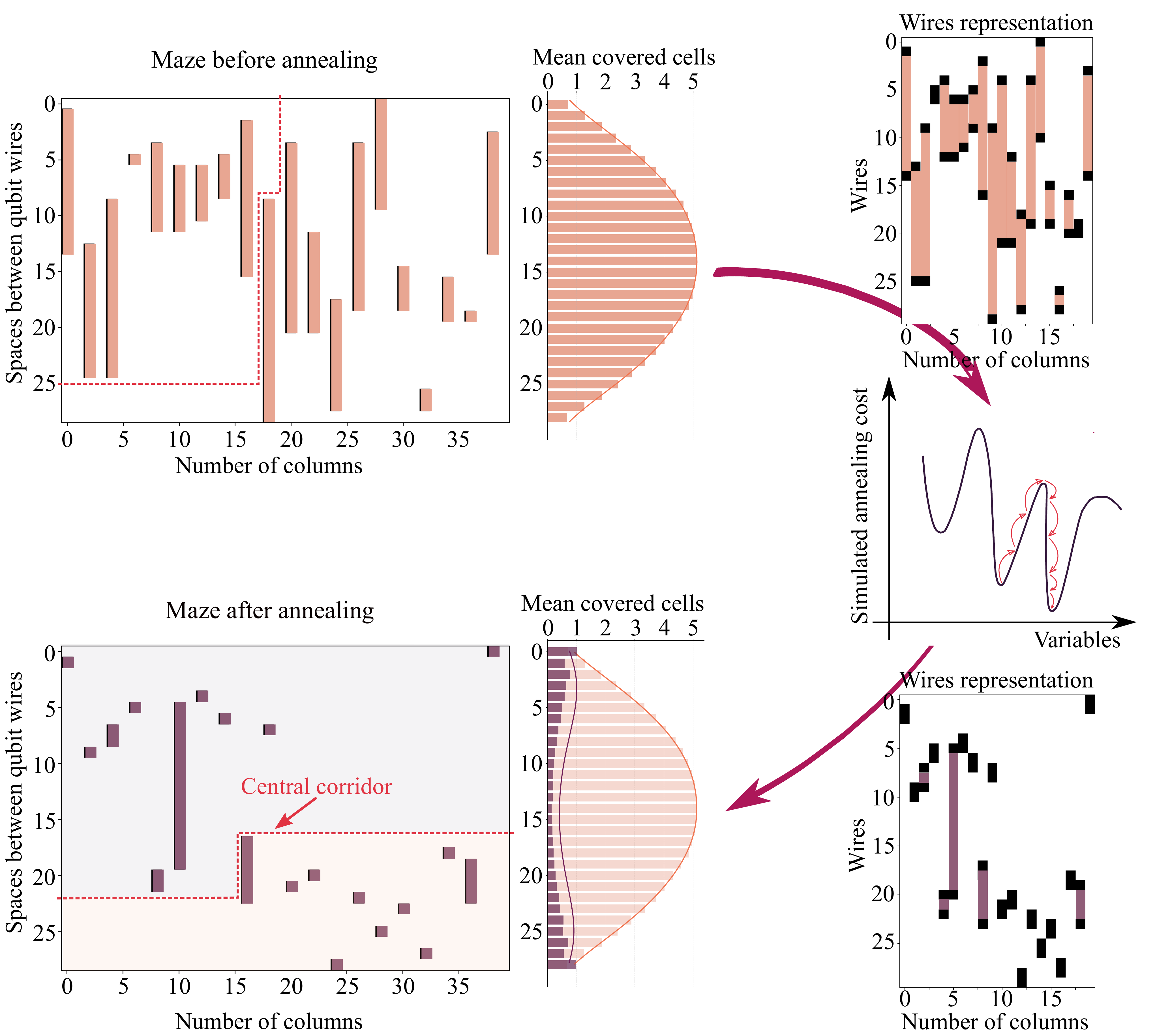}
    \caption{\raggedright The maze layout optimization using simulated annealing on the wire representation of the circuit. 
    The example illustrates a circuit of width $N_q=30$ and depth $N_c=20$. 
    The initial configuration (top left) features a high-density central region that prevents the formation of a balanced partition by the cut represented by the red dashed line. 
    This is confirmed by the cell distribution (on the immediate right of the maze), where $10000$ simulations show a characteristic peak in the central region. 
    Applying the SA algorithm to the intermediate wire representation, a permutation is found that clusters the CNOT interactions toward the up and bottom periphery. 
    The resulting optimized maze (bottom left) features a low dense central corridor and two partitions balanced in terms of both qubit density and CNOT distribution, significantly improving traversal for an heuristic agent.}
    \label{fig:simulated annealing}
\end{figure*}
Indeed, if a cutting path curves into a hump, operations assigned to one partition can appear before, during, and after operations assigned to the other at the hump.
Such a cut would therefore introduce a chronological interleaving between the two subcircuits created by the partition, preventing independent resynthesis.

In the following, we will exploit the complementary strengths of the $\mathcal{W}$ and $\mathcal{M}$ representations, using the wire representation to manipulate the ordering of qubits and the maze representation to construct path-based feasible circuit cuts.

\section{Optimization of qubit ordering for maze geometry}
\label{section:qubit reordering}

In this Section, we show that, since the distribution of walls in the maze $\mathcal{M}$ peaks at the center, it is necessary to manipulate the order of the qubit wires in $\mathcal{W}$. 
As depicted in Figure \ref{fig:simulated annealing}, this manipulation creates a central region free of walls, henceforth referred to as the central corridor, which favors paths passing through the center of $\mathcal{M}$.
Where possible, this leads to a bipartite non trivial partition of the walls. 
To this end, we define a cost function which is minimized by using simulated annealing (SA) \cite{kirkpatrick1983optimization}. 
We then examine the outcome of this approach in two ways.
We first examine the evolution of the maze walls distribution under the influence of SA, which transforms such a distribution from a single-peaked to a double-peaked one.
Next, we study both the centrality measure of the graph associated with the circuit and the corresponding information-flow dynamics, showing that SA favors the emergence of weakly coupled qubit regions when the number of CNOT gates is low, whereas this separation gradually disappears as the number of CNOT gates increases.

\subsection{Asymptotic distribution of covered cells}

As said, the quantum circuits we consider are constructed such that the CNOT gates are randomly distributed among the qubits. 
By placing the qubits on a circle, each CNOT gate is represented by an edge connecting two boundary points.
Averaging over many circuits, all pairs of qubits are connected with approximately equal probability, so that the resulting graph is, on average, fully connected.
After cutting the circle between two randomly chosen adjacent qubits and flattening it, a vertical line intersects precisely those CNOT edges whose endpoints lie on opposite sides of the cut.
Therefore, the number of crossed edges determines the number of covered cells in the corresponding $\mathcal{M}$ row.
This fact explains why the distribution of the walls is not uniform. 
Indeed, the number of intersections is smaller near the edges and larger in the center, bringing the profile into the continuum limit.

More precisely, if the maze has $N_{\text{MR}}$ rows, indexed by $k = 0, ..., N_{\text{MR}} - 1$, then the average number of covered cells in row $k$ is 
\begin{equation}
    a_k = -k^2 + (N_{\text{MR}} - 1)k + N_{\text{MR}}
\end{equation}
After normalisation, $a_k$ defines a discrete probability distribution over the rows. 
Introducing the rescaled variable $y = k/N_{\text{MR}}$, with $n = N_{\text{MR}} - 1$, and considering the continuous limit $N_{\text{MR}}\to\infty$, we find that the corresponding density converges to 
\begin{equation}
    p(y) = 6y(1 - y)\,\,\,\,\text{for}\,\, y \in [0, 1]
\end{equation}
with $\int_0^1 p(y) dy = 1.$ 
Hence, the wall distribution approaches a $\text{Beta}(y,2,2)$ law in the large-size limit.
The complete derivation of this result is provided in Appendix \ref{appendix:p_before_annealing}.

\subsection{Wire permutation via simulated annealing}

As shown in the previous section, the wall distribution of the maze $\mathcal{M}$ is, on average, peaked around the center and approaches a $\mathrm{Beta}(y,2,2)$ profile in the continuous limit.
As a consequence, the central region of the maze is typically the most obstructed one, making horizontal traversal unlikely and therefore reducing the possibility of obtaining balanced circuit partitions.

A natural way to reduce this obstruction is to reorder the qubit wires.
However, the rows of the maze $\mathcal{M}$ cannot be permuted directly, since they represent the physical spaces between adjacent qubit wires.
Instead, we permute the rows of the wire representation $\mathcal{W}$, where each row corresponds to a qubit wire.
Once a new ordering of $\mathcal{W}$ is selected, the corresponding maze $\mathcal{M}$ is reconstructed from it.
In this way, the geometry of the maze is modified through a physically consistent permutation of the qubit layout.

The goal of the permutation is to move CNOT interactions away from the center of the grid and towards the upper and lower regions.
In view of creating a cost function, we associate with each wire configuration an auxiliary binary matrix $ x_{ij} \in {0,1}^{N_{WR}\times N_{WC}},$ which records the full vertical span of each CNOT gate.
More precisely, for the $k$-th CNOT acting on qubits $c_k$ and $t_k$, all entries between the two corresponding wire positions are set to one:
$$ x_{ik}=1 \quad \text{for} \quad \min(c_k,t_k)\leq i \leq \max(c_k,t_k), $$
and zero otherwise.
Thus, while $\mathcal{W}$ marks only the two wires involved in a CNOT gate, the matrix $x$ also marks all wires crossed by that interaction.
We then assign to each wire layout the cost
\begin{align}
\mathcal{C} = \sum_{i=1}^{N_{WR}} \sum_{j=1}^{N_{WC}} \omega_i x_{ij},
\label{eq:cost_SA_def}
\end{align}
where
\begin{align}
\omega_i = \exp\left[-\frac{1}{2N_{WR}}\left(i-\frac{N_{WR}}{2}\right)^2\right].
\label{eq:omega_i_weights}
\end{align}
The weights $\omega_i$ are largest near the center of the grid and decrease toward the boundaries.
Therefore, a CNOT whose vertical span crosses the central region gives a larger contribution to $\mathcal{C}$ than a CNOT localized near the edges.
Minimizing $\mathcal{C}$ consequently favors wire orderings in which the CNOT spans are pushed away from the center, creating a less obstructed central corridor in the maze.

We minimize this cost by simulated annealing (SA) over the space of wire permutations.
At each step, the algorithm proposes a new ordering of the rows of $\mathcal{W}$, reconstructs the corresponding span matrix $x$, and accepts or rejects the move according to the SA rule.
The procedure is run for a fixed number of iterations, $N_{SA}=100000$.
The optimized wire representation is finally mapped back into the maze representation, producing a layout in which the interaction regions are more clearly separated and the central corridor is more accessible to the traversal heuristic, as illustrated in Figure~\ref{fig:simulated annealing}.

\subsection{Evolution of the ansatz for the SA-dependent maze distribution}

\begin{figure*}[t]
    \centering
    \includegraphics[width=\textwidth]{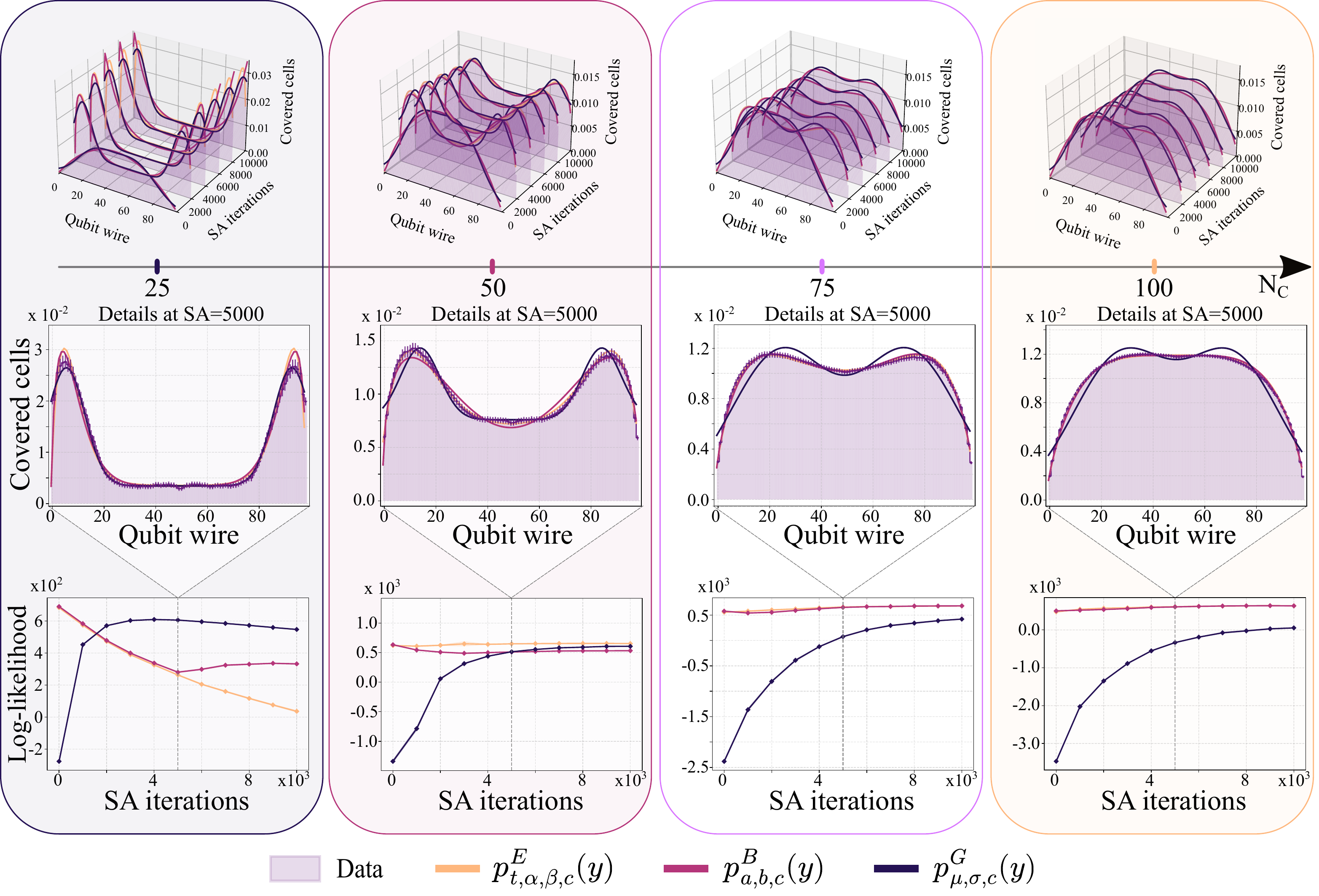}
    \caption{\raggedright The evolution of the average distribution of covered cells under simulated annealing for $N_q=100$ and $N_c=25, 50, 75, 100$. 
    The top panels show the distribution of qubits across the wires as a function of SA iterations up to $N_{\mathrm{SA}}=10^4$, while the middle panels compare the empirical profile at $N_{\mathrm{SA}}=5000$ with the three normalized fits $p^E_{t,\alpha,\ beta,c}(y)$, $p^B_{a,b,c}(y)$, and $p^G_{\mu,\sigma,c}(y)$. 
    The bottom panels show the corresponding Gaussian log-likelihoods as a function of SA iterations, highlighting the ansatz with the best performance in each phase of evolution. 
    For $N_c=25$, the the distribution in the late phase is best described by the symmetric Gaussian mixture assumption, while for
    $N_c=50$ and $N_c=75$, the effective model $p^E_{t,\alpha,\beta,c}(y)$ provides the best overall agreement, although the difference for certain steps is minimal.
    For $N_c=100$, the effective model and the beta-type model, $p^E_{t,\alpha,\beta,c}(y)$ and $p^B_{a,b,c}(y)$, remain nearly indistinguishable across all SA iterations and outperform the Gaussian mixture assumption.}
    \label{fig:distribution fit}
\end{figure*}

The average $p(y)$ distribution evolves during the SA process as the underlying layout of the maze rows is progressively updated. 
To track how the distribution evolves over the SA iterations, we fit the empirical data using three parametric models selected to accurately capture the symmetric, often bimodal structure observed in the circuit samples.
Specifically, we consider a symmetric mixture of Beta distributions $p^{B}_{a,b,c}$, a pair of symmetrically placed Gaussian components $ p^{G}_{\mu,\sigma,c}$, and a boundary-weighted density with a localized exponential modulation $p^E_{t,\alpha,\beta,c}$.

For a circuit with $N_{q}$ qubits, the discrete row index $i=0,\dots,N_{q}-2$ is mapped to the rescaled coordinate

\begin{align}
    u_i=\frac{i}{N_{q}-1}\in[0,1].
\end{align}

The empirical distribution is then normalized and compared with the fitted candidates.

The first ansatz is a symmetric combination of Beta distributions with an additional background term:
\begin{align}
    \displaystyle p^{B}_{a,b,c}(y)\propto B(y;a,b) + B(y;b,a) + c
\end{align}
where $B(\cdot\,;a,b)$ denotes the Beta density with parameters $a$ and $b$. 
Such form is symmetric with respect to $y=\tfrac12$ and naturally produces two peaks of equal height, one on each side of the center. 
The additive term $c$ allows for a finite background contribution and is useful because the central region is never completely empty. 
The ansatz is flexible enough to interpolate both bimodal profiles symmetric about the center and the single-peak distribution before the SA effect, namely $p_0(y) = \text{Beta}(y,2,2)$.

The second ansatz models the distribution as the sum of two Gaussians positioned symmetrically around the midpoint, plus a constant offset:
\begin{align} 
\displaystyle p^{G}_{\mu,\sigma,c}(y)\propto e^{-\frac{(y-\mu)^2}{2\sigma^2}} + e^{-\frac{(y-(1-\mu))^2}{2\sigma^2}} +c
\end{align}
Here, $\mu\in[0,\tfrac12]$ determines the position of the left peak, while the second peak is fixed at $1-\mu$ to preserve symmetry. 
The parameter $\sigma$ controls the width of the two peaks, and $c$ accounts for a non-zero vertical offset. 

For the third ansatz, we model the effect of the SA algorithm on the wall distribution, using a coarse-graining procedure and an effective potential $V_{\text{eff}}(y)$, which should model the separation of the original Beta distribution $p_0(y)$ into two peaked distributions that are symmetric around $y=1/2$. 
A suitable ansatz for the effective potential can be inferred from the shape of the $\omega_i$ weights in Eq.~\eqref{eq:omega_i_weights}.
In the continuous limit the weights approximate a continuous function $\omega(y)$ that has the required properties to model the effective potential (see also Appendix \ref{sec:effective_potential} for details), so that we can set:
\begin{align}
    V_{\text{eff}}(y) \approx N_q \omega(y),
\end{align}
and 
\begin{align}
    \omega(y) = \exp{- \frac{1}{2 \sigma^2 N_q^2}\left(y-\frac{1}{2}\right)^2 }.
\end{align}
The probability distribution after a sufficiently long annealing can be approximated as:
\begin{align}\label{eq:fit_exp_explanation}
    p^*_T(y) = \frac{1}{Z^*} p_0(y) \exp{-\frac{N_q \omega(y)}{T}},
\end{align}
where $Z^*$ is the normalization constant of $p^*_T(y)$.
Due to boundary effects, we also have to account for offsets in the probability distribution that are not included in Eq.~\eqref{eq:fit_exp_explanation}, so that our final model for the probability distribution post-SA becomes
\begin{align}
    \Tilde{p}^*_T(y) = b(y) + M p^*_T(y)
\end{align}
where $b: \mathbb{R} \mapsto \mathbb{R}_{>0}$ is a boundaries-correction function that forces the distribution to have non-zero values at the boundaries of $[0,1]$, and
\begin{equation}
     M = 1 - \int_0^1 b(y) dy \label{eq:M_b}
\end{equation}
is a normalization constant. 
A more detailed discussion is provided in \ref{eq:M_b}.

By combining the theoretical concepts outlined above, we define the fitting curve
\begin{equation}\begin{split}
    p^E_{t,\alpha,\beta,c}(y)\propto \,c\, +\,\beta\big(e^{-y} + e^{y-1}\big)\, +\quad\quad\quad\,\,\, \\
    \,\, + \, y(1-y)\left[ \exp\!\left( -\frac{1}{t}e^{-\alpha\left(y-\frac12\right)^2} \right) \right].
\end{split}
\end{equation}
The first term is the offset $c$, while the second term is a weighted sum of exponential functions which ensure that at the interval boundaries the distribution does not converge to zero.
The third term in the end is a deformation of the profile $y(1-y)$ by an exponential factor that selectively suppresses the probability near the center.
This form allows the weight near the center to decrease as the bimodal structure develops. 
The parameter $\alpha$ represents a value proportional to the inverse variance and determines the degree of localization of the central suppression around $y=\tfrac12$.
The parameter $t$ plays a role analogous to an effective temperature and therefore controls the overall strength of the suppression term.

In all three cases, the output of the model is properly normalized.

\subsection{Fitting procedure and model selection}

Every $N_{SA} = 1000$, the parameters of the three approaches are optimized and compared with each other and the data.
For each snapshot, the empirical distribution is reconstructed by averaging over $10000$ random quantum circuits.
The uncertainty associated with each bin of the resulting mean distribution is then estimated and propagated throughout the entire fitting procedure.

Let $y_i$ be the mean value observed at $u_i$, and let $\sigma_i$ be the associated standard error. 
For a model prediction $f(u_i;\theta)$, the fit is performed by minimizing the chi-square function
\begin{equation}
    \chi^2(\theta)=\sum_i r_i(\theta)^2=\sum_i \left[\frac{f(u_i;\theta)-y_i}{\sigma_i}\right]^2.
\end{equation}

The parameters of each hypothesis are estimated through a two-stage optimization procedure using routines provided by the \texttt{SciPy} library.
First, a global exploration of the parameter space is performed using differential evolution \cite{storn1997differential} to reduce sensitivity to initialization and avoid suboptimal local minima.
The best candidate identified in this global step is then used as the initial condition for a local refinement based on \texttt{scipy.optimize.least\_squares}.
Since the local solver is used with a linear loss function, this second step corresponds to a standard weighted nonlinear least squares minimization of the residuals $r_i(\theta)$, and therefore to a direct local minimization of the chi-square objective function.
In this way, the optimization combines the robustness of a global search with the efficiency and precision of local least squares refinement close to the optimum.
Once the optimal parameter set $\hat{\theta}$ for each ansatz is obtained, the estimated distribution is compared to the empirical one using the Gaussian log-likelihood
$$
\log L = -\frac12\sum_i\left[\log(2\pi \sigma_i^2) + \left(\frac{y_i-f(u_i;\hat{\theta})}{\sigma_i}\right)^2\right]
$$
to quantify the agreement between the estimated ansatz and the empirical distribution.

The results are shown in Figure~\ref{fig:distribution fit} for the case $N_{c}=\,25,\, 50, \,75, \,100$ and $N_q=100$, up to $N_{SA}=10000$. 
The best ansatz which describes the data at the end of SA action depends on $N_{c}$. 
In the $N_{c}=25$ case, the walls are perfectly divided into two clusters.
Between $N_{SA}=0$ and $N_{SA}=1500$, the best ansatz representing the transition of the distribution from a single peak to a bimodal one is the beta combination $p_{a,b,c}^B$. 
Such a result is predictable because the original distribution of the $N_{SA}=0$ is actually a $\text{Beta}(y,2,2)$.
Beyond $N_{SA}=1500$, the log-likelihood indicates that the symmetric Gaussian combination $p_{\mu,\sigma,c}^G$ provides the best approximation to the distribution. 
Once the two clusters have been clearly identified and separated, there are no more CNOT threads connecting the first cluster to the second. 
When this happens, the two groups become effectively independent and their distributions are well approximated by two Gaussian profiles.

For $N_c=50$ and $N_c=75$, $p^E_{t,\alpha,\beta,c}(y)$ provides the best overall approximation, although the differences with $p^B_{a,b,c}(y)$ remain small for many SA iterations.
For $N_c=100$, $p^E_{t,\alpha,\beta,c}(y)$ and $p^B_{a,b,c}(y)$ are almost indistinguishable along the entire SA trajectory, and both outperform the Gaussian mixture ansatz.
This behavior can be understood by considering that, for $N_c=50, 75, 100$, the CNOT gates cannot be clearly separated into two distinct groups.
Consequently, a weakly bimodal profile provides a more appropriate description, which explains why $p^E$ and $p^B$ models produce the best agreement with the data.
In such regimes, we expect that a heuristic traversing the maze while attempting to balance the two partitions cannot divide the maze horizontally, but that a more complex path must be found.

\subsection{Simulated annealing effect on qubits importance}
\label{sec:centralities}

To better understand the effect of the SA algorithm, we develop a graph-based approach.
We construct a directed graph $\mathcal{G}(\mathcal{V},\mathcal{E})$ associated with the circuit as follows. (i) For each CNOT gate in the circuit we create two nodes, one placed on the control wire and one on the target wire; thus a circuit with $N_c$ CNOT gates yields $|V| = 2N_c$ nodes. (ii) For every CNOT gate, the two nodes corresponding to its control and target are joined by a bidirectional edge. (iii) Along each qubit wire, the nodes corresponding to successive CNOT events on that wire are joined by directed edges that follow the temporal order of the circuit. The bidirectional CNOT edges encode the entangling interactions, while the directed wire edges encode the causal, time-ordered structure of the circuit. This is the same graph used for the coarse-grained consensus analysis, and its layered, time-expanded structure is described in detail in Appendix \ref{app:graph_cons}.\\

\begin{figure*}
    \centering
    \includegraphics[width=\textwidth]{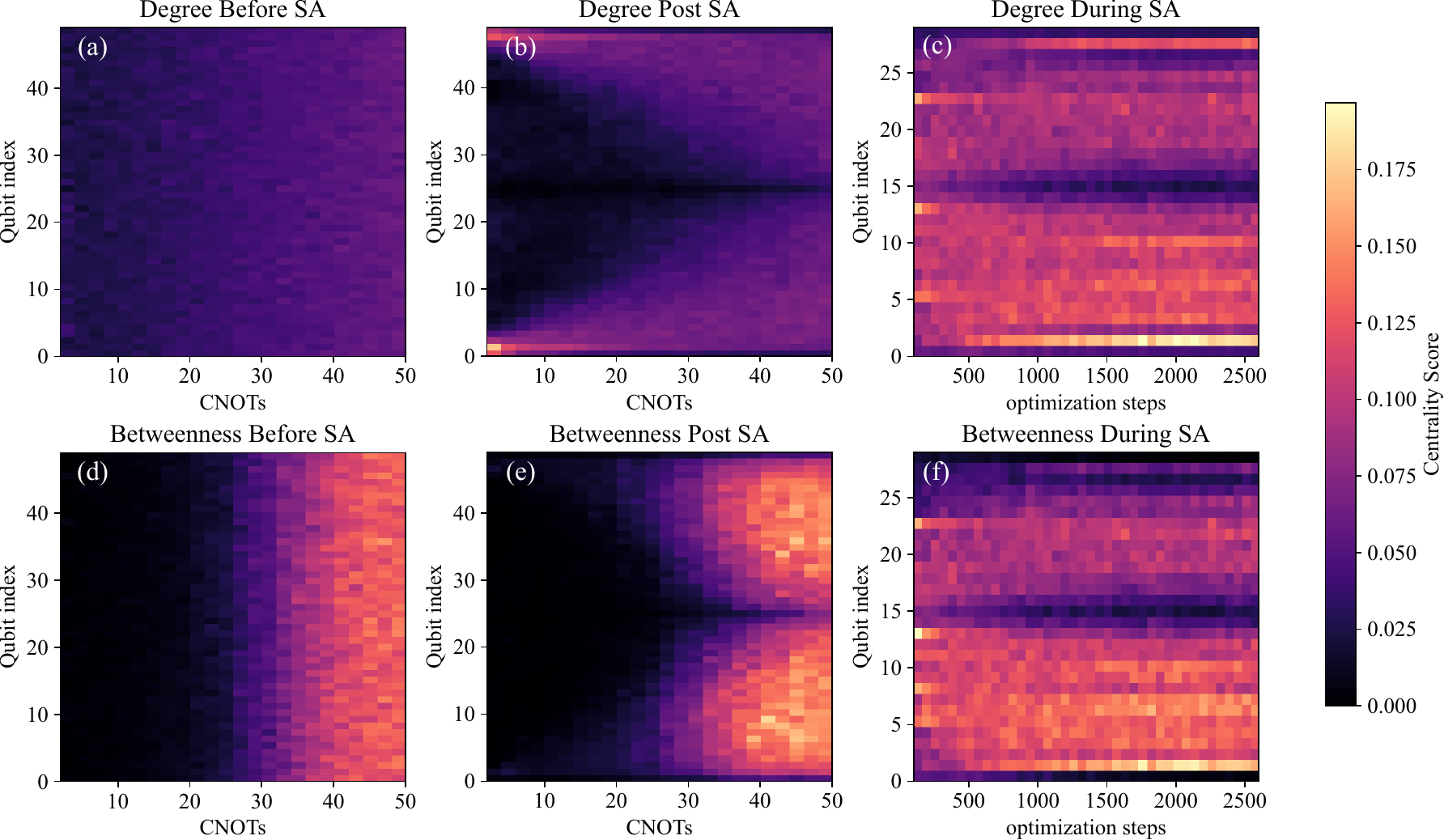}
    \caption{\raggedright
        Panels~(a) and~(d) show the degree and betweenness centrality, respectively, before the SA optimization, for an ensemble of 500 random quantum circuits with up to $N_q = 50$ qubits and $N_c = 50$ CNOT gates. The $x$-axis represents the number of CNOT gates and the $y$-axis the qubit index, with each cell showing the centrality averaged over the ensemble. Before optimization, both 
        centralities are approximately uniform across qubits for any fixed number of CNOTs, reflecting the absence of any preferred interaction structure in randomly generated circuits. Panels~(b) and~(e) show the same quantities after SA optimization. A markedly different structure emerges: for low to intermediate CNOT counts, the optimized circuits display a pronounced inhomogeneity, 
        with qubits at the edges of the register acquiring significantly higher centrality than those in the center. This corresponds to the formation of the central corridor, as the annealer pushes interactions toward the boundaries of the register following the cost function. 
        As the number of CNOTs increases toward $N_c \approx 30$, the corridor progressively narrows, reflecting the reduced degrees of freedom available to the optimizer in denser circuits. Panels~(c) and~(f) show the evolution of degree and betweenness centrality during the SA training for a fixed circuit with $N_c = 30$ CNOT gates, as a function of optimization steps ($x$-axis). The inhomogeneous structure develops progressively, with the boundary qubits acquiring their high centrality scores as the number of optimization steps increases, clearly illustrating how the SA procedure actively reshapes the interaction graph over the course of the optimization.
    }
    \label{fig:Centralities}
\end{figure*}

We explicitly use directed edges to capture the causal structure of the circuit. We model the entangling edges as bidirectional. Because single-qubit gates are absent from both the maze and wire representations, the entangling edge is specified only up to local Clifford operations; within this representation a CNOT and a CZ which is related by $\text{CNOT} = (\text{I}\otimes \text{H})\text{CZ}(\text{I}\otimes \text{H})$ interchangeably. Since CZ acts symmetrically on its two qubits, we treat the entangling edge as undirected, with neither qubit designated as a privileged source of influence. The time-ordered wire edges remain directed, preserving the causal structure of the circuit. Time-directed edges connect successive CNOT events on the same qubit wire, preserving the temporal ordering of the circuit.
By using this construction we can study different aspects of SA from two complementary perspectives, namely the graph topology and the information flow.

First we study the behaviour of topology using as metrics two node centralities. 
We take into account two centralities, the degree centrality (Eq. \ref{eq:deg}) and the betweeness centrality (Eq. \ref{eq:betw}) \cite{barthelemy2004betweenness}.
The first one is defined as:
\begin{equation}\label{eq:deg}
    C_D(i) = \deg(i) = \sum_{j \in V} A_{ij},
\end{equation}

where $A$ is the adjacency matrix of the related graph $\mathcal{G}$. We can interpret a qubit with a high degree centrality as a qubit that participates in many CNOT interactions.\\
The second centrality measure we are taking into account is the betweeness centrality defined as 
\begin{equation}\label{eq:betw}
    C_B(i) = \sum_{s \neq i \neq t}\frac{\sigma_{st}(i)}{\sigma_{st}},
\end{equation}

where $\sigma_{st}$ is the total number of shortest paths between nodes $s$ and $t$, $\sigma_{st}(i)$ is the number of those paths that pass through node $i$.

The centrality of a qubit wire j is defined as the average centrality over all event-nodes ($i$) lying on that wire:
\begin{equation}
    C_{D/B}(j) = { \frac{\sum_{i\in \mathcal{V}} C_{D/B}(i)\chi_j(i)}{\sum_{i\in \mathcal{V}} \chi_j(i)}},
\end{equation}

where $\chi_j(i)$ represents the indicator function 
\begin{equation}
    \chi_j(i) = \begin{cases}
    0 \quad i \not\in \mathcal{V}_j\\
    1 \quad i \in \mathcal{V}_j
\end{cases} ,
\end{equation}
and $V_j \subseteq \mathcal{V}$ denotes the set of nodes associated with qubit wire $j$. In words, $C_{D/B}(j)$ is the mean degree (resp. betweenness) centrality of the CNOT events occurring on wire j, and quantifies how strongly that qubit participates in or mediates the entangling interactions of the circuit. We simulate $500$ random circuits for each parameter setting, fixing the number of qubits to $N_q=50$ and varying the number of CNOT gates in the range $N_c\in[0,50]$. 
For each circuit instance, both centrality measures are evaluated.

In addition, we investigate the effect of the SA training process as a function of the optimization time. 
The algorithm is trained for a number of optimization steps ranging from $100$ to $2600$, and the state of the optimization is sampled every $40$ steps.

Figure~\ref{fig:Centralities} shows the effect of the simulated annealing optimization on the circuit interaction graph. 
In non-optimized circuits (panels (a),(d)), both degree and betweenness centralities are relatively homogeneous across qubits, indicating that randomly generated circuits distribute two-qubit gates rather uniformly over the qubit register.

After optimization (panels (b),(e)), a markedly different structure emerges. 
For small to intermediate numbers of CNOT gates $(0-30)$, the optimized circuits display a pronounced inhomogeneity in both centrality measures, with a subset of qubits -- predominantly those located toward the edges of the register --acquiring a larger topological relevance. 
This indicates that the simulated annealing procedure actively restructures the circuit to concentrate interactions into specific qubit lanes trying to avoid the central ones.

As the number of CNOT gates is further increased, this separation becomes progressively less effective. 
Both degree and betweenness centralities tend again toward a more homogeneous distribution, suggesting that, at higher circuit density, the simulated annealing algorithm struggles to maintain a clear topological differentiation among qubits. 
This behavior points to an intrinsic limitation of the optimization in disentangling highly connected circuits and highlights a transition from structured to saturated interaction graphs.

Panels (c) and (f) track the same quantities along the SA trajectory for a fixed circuit with $Nc=30$: the boundary–edge inhomogeneity builds up progressively and saturates, confirming that the restructuring is driven by the optimization dynamics rather than being a property of the initial random circuit.

\subsection{Connectivity structure and consensus dynamics}

The centrality analysis identifies which qubits dominate the entangling interactions, but it does not by itself establish that the two candidate partitions are separable, since qubits in different partitions may still be linked by CNOT gates crossing the intended cut. To assess this directly, we introduce a coarse-grained consensus dynamics that measures how far a given circuit departs from an idealized, perfectly partitioned one.
The idea is to take the ideal bipartition induced by the maze cut as a reference configuration and let the interaction structure of the circuit act on it. The conductance analysis -- see Appendix \ref{app:Conductance} -- provides the quantitative graph-theoretic measure of inter-partition coupling; the consensus dynamics offers a complementary dynamical picture of the same connectivity, showing how strongly the actual circuit deviates from that ideal separation. The two analyses probe the same underlying connectivity and are therefore consistent by construction rather than independent. In this section we present the consensus results as a dynamical illustration of the connectivity regimes.

The directed graph representation previously introduced can also be interpreted from the perspective of classical consensus models on networks. 
In particular, the dynamics implemented in our numerical experiments corresponds to a variant of the \emph{DeGroot consensus model} \cite{degroot1974reaching}, widely used in opinion dynamics.

In our simulations we associate a scalar state $x_i(t)$ to each node of the circuit graph and evolve it according to the update rule
\begin{equation}
x_i(t+1) = (1-\epsilon)x_i(t) + \epsilon \frac{1}{|\mathcal{N}_i^{\mathrm{out}}|}
\sum_{j\in \mathcal{N}_i^{\mathrm{out}}} x_j(t),
\label{eq:first_cons}
\end{equation}
where $\mathcal{N}_i^{\mathrm{out}}$ denotes the set of outgoing neighbors of node $i$. 
This latter update rule corresponds to a local averaging process in which each node retains a fraction $(1-\epsilon)$ of its own state while incorporating information from its neighbors.

This dynamics -- see Eq.~\eqref{eq:first_cons} -- can be rewritten in the standard DeGroot form:
\begin{equation}
    x_i(t+1) = \sum_j W_{ij} x_j(t).
\end{equation}
In our case the weights are given by
\begin{equation}
\begin{cases}
W_{ii}& = 1-\epsilon, \\
W_{ij}&= \frac{\epsilon}{|\mathcal{N}_i^{\mathrm{out}}|}, 
\qquad j \in \mathcal{N}_i^{\mathrm{out}},
\end{cases}
\end{equation}
so that $\sum_j W_{ij} = 1.$

The presence of the self-weight $W_{ii}$ makes the process a \emph{lazy} consensus dynamics, meaning that each node partially retains its own value at every step. 
In the language of stochastic processes this is equivalent to a random walk on the graph with self-loops, which improves stability and convergence properties in directed networks.

In our setting the influence network is directly derived from the circuit structure as shown in Section \ref{sec:centralities}.

We recall that nodes correspond to CNOT events, and directed edges represent causal relationships between successive gate operations. Time-directed edges connect consecutive events on the same qubit wire, encoding the natural temporal ordering of the circuit. For edges associated with CNOT gates, we adopt a bidirectional convention, motivated by the symmetric nature of entanglement generation: a CNOT creates correlations between control and target qubits that are mutually dependent, so neither qubit can be considered the sole source of influence on the other.
Running the DeGroot consensus process on this graph provides a coarse-grained probe of connectivity between circuit regions. The consensus dynamics spreads a scalar quantity along the same interaction structure defined by the CNOT gates, without tracking individual quantum operators. In this interpretation, the long-time behavior of the process reflects how strongly different regions of the circuit are coupled through their shared entangling interactions: weakly coupled regions preserve the initial separation between the two populations, while strongly coupled ones mix them.

To probe this structure we initialize the node states sampling their values from a uniform distributions with opposite signs: in the upper $x_{i_{up}} \sim \mathcal{U}(0,1)$ and lower $x_{i_{bot}} \sim \mathcal{U}(0,-1)$ halves of the circuit and evolve the consensus dynamics for a fixed number of iterations (5000).

We quantify the separation between the two populations by fitting a two component Gaussian Mixture Model (GMM) -- see Appendix \ref{app:GMM} -- and computing the overlap area between the two Gaussians:
\begin{equation}
A = \int \min\!\left[ p_1(x), p_2(x)\right] dx ,
\end{equation}
where $p_1$ and $p_2$ denote the two Gaussians.

A small value of A indicates a strong separation between the two halves of the circuit, while larger values corresponds to stronger mixing between them.

As the circuit becomes denser (increasing ratio of $\frac{N_c}{N_q}$), the propagation graph becomes more strongly connected and the consensus dynamics mixes more efficiently across qubits, leading to an increase in the overlap area. 
Conversely, when the simulated annealing procedure pushes interactions toward the boundaries of the grid, the graph develops two weakly coupled regions and the DeGroot dynamics separates into two clusters. 
In this regime the overlap area becomes small, reflecting the emergence of a corridor-like structure that restricts information flow across the circuit.

\begin{figure}
    \centering
    \includegraphics[width=\linewidth]{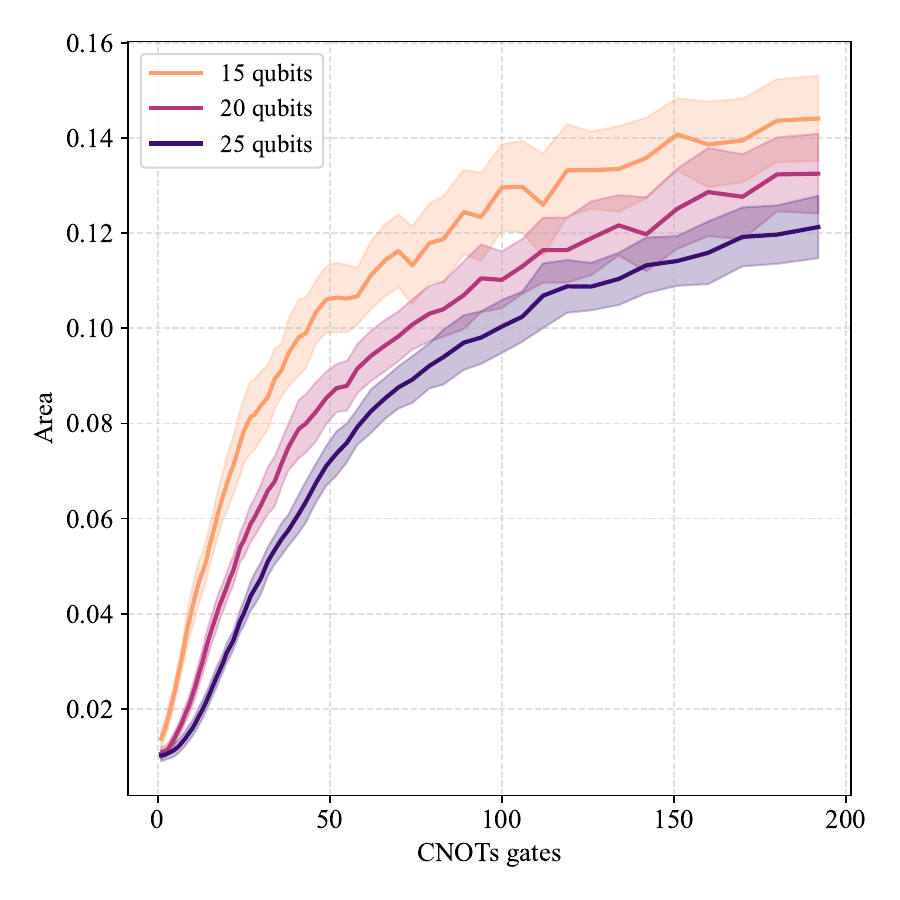}
    \caption{\raggedright
    The overlap area $A$ between the two Gaussian components of the GMM fitted to the final consensus values, shown as a function of the number 
    of CNOT gates for circuits with $N_q = 15$, $20$, and $25$ qubits. The $x$-axis represents the number of CNOT gates and the $y$-axis the overlap area $A$, where small values indicate strong separation between 
    the two partitions and larger values correspond to stronger mixing. Shaded regions indicate the standard deviation over 500 circuit realizations. For all system sizes, $A$ remains close to zero at low circuit densities, confirming that the SA algorithm successfully separates the two partitions in the sparse regime. As the number of 
    CNOT gates increases, the overlap grows monotonically for all system sizes, reflecting the progressive breakdown of the bipartite structure. 
    Smaller systems reach larger overlap values at lower CNOT counts, consistent with the reduced degrees of freedom available to the optimizer 
    in denser circuits relative to the system size.
}
    \label{fig:area-SA}
\end{figure}

As expected, this behavior is clearly observed in Figure~\ref{fig:area-SA}. 
The overlap area increases monotonically with the number of CNOT gates for all system sizes, indicating progressively stronger mixing across the circuit. The three curves are well separated, with smaller systems reaching higher 
overlap values at lower CNOT counts, reflecting the fact that a fixed number of entangling gates has a proportionally larger impact on smaller registers. 
For $N_q = 15$ qubits, the overlap grows more steeply and reaches larger values at intermediate CNOT counts, consistent with the reduced number of 
degrees of freedom available to the optimizer. For $N_q = 20$ and $N_q = 25$, the growth is more gradual over the explored range, suggesting that residual 
separability between the two halves of the circuit persists up to higher circuit densities. In all cases, the overlap area remains small for low CNOT counts, confirming that the SA algorithm successfully separates the two 
partitions in the sparse regime.

The same trend is captured quantitatively by the conductance of the interaction graph -- see Appendix \ref{app:Conductance} --, which measures the coupling between the two ideal partitions directly from the graph structure.

\section{Heuristics for traversing the maze and the emergence of a phase transition}
\label{section:phase transition}

In this Section, we introduce a maze-cutting heuristic and show that, when we apply this technique, quantum circuits exhibit two distinct regimes, characterized by a phase transition in the minimum mean path length as a function of the number of qubits and the circuit depth.

\begin{figure*}[t]
    \centering\includegraphics[width=\textwidth]{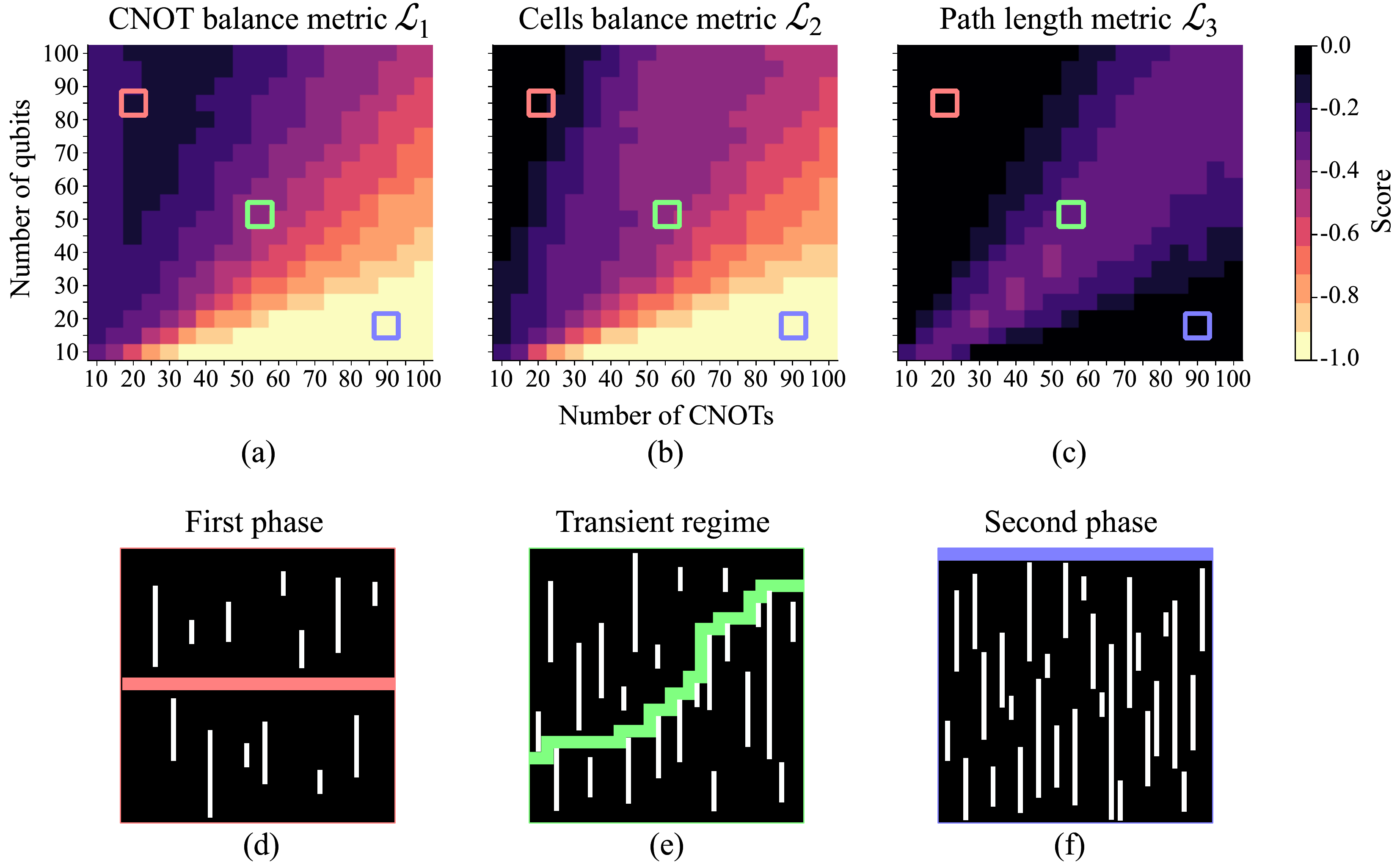}
    \caption{\raggedright Score obtained by the best performing agent among all candidates at $\alpha=1, \beta=1, \gamma=2$ for each evaluation metric considered, based on the structural parameters of the maze, i.e., the number of qubits (which determines the number of rows), and the depth of the CNOT circuit (which sets the number of columns).
    (a) The first metric penalizes imbalances in the number of CNOT gates assigned to the two partitions, favoring cuts that distribute the CNOTs as evenly as possible. 
    (b) The second metric penalizes uneven division of the maze itself, encouraging an equal number of maze cells in each partition. 
    (c) The third metric evaluates the length of the path built by the agent traversing the maze. Such a metric penalizes deviations in path length from the ideal progression per column, thus favoring horizontal paths. 
    Each contribution is normalized to ensure comparable influence, and their weighted combination determines the final score used to select the optimal cut.
    Based on the values of these three metrics, three different regimes emerge, shown at the top of the figure. 
    (d) In the first phase, (also indicated by the pink square in each figure below), the best agent traverses the maze horizontally, keeping the number of CNOTs and cells almost in balance. 
    (f) The second regime (also marked by the blue square in each figure below), where no division of the maze is possible and even the best agent is forced to follow a trajectory outside the physical maze.
    (e) The region in the center is a transitional regime between the two previous phases and is (indicated by the green square at the center of each figure below).
    The trajectories are sometimes horizontal and sometimes vertical, creating a succession of steps upwards and to the right or downwards and to the right. 
    }
    \label{fig:maze metrics}
\end{figure*}

After performing the SA optimization, the new maze is ready to be cut.
Among the various possible options, the most feasible is to use a heuristic method that addresses two fundamental problems in a simple and efficient way.
The first is how to choose the starting cell on the leftmost column from which to begin the cut.
The second is to consider only heuristics that create a monotone maze cutting path, avoiding the complications associated with humps, as explained in Section \ref{section:wire&maze}.
We tackle both of these problems by implementing the simplest, fastest, and easiest-to-calculate heuristic.
For each cell in the first column of the maze, we initialize two deterministic agents.
The first agent follows an upward policy.
It can only move upward when it encounters a wall in the maze, or to the right when the next cell is empty.
The second agent follows a downward policy, meaning it can only move downward when it encounters a wall, or to the right otherwise.
This constrained grid dynamics for both agents starts on each cell in the first column and ensures that the resulting maze path is monotone.
The total number of agents depends linearly on the number of qubits in the circuit as $2(N_{q}-1)$, and each agent is independent of the others, meaning we can launch them in parallel. 
Note that the path cost of each agent is always less than or equal to the total length of the maze diagonal, which is $N_{MC} + N_{q} - 1$.
To select which candidate cut-path is the most suitable, we implement the following weighted cost function:
\begin{equation}\label{eq:l_score_tot}
  \mathcal{L}_{tot} = \alpha\,\mathcal{L}_1+ \beta\,\mathcal{L}_2 + \gamma\,\mathcal{L}_3,  
\end{equation}
where the first term:
\begin{equation}
    \displaystyle \mathcal{L}_1 = -2\,\frac{|\Delta_{CNOTs}|}{N_{MC}-1}\,\,\,\,\in\,[-1,0],
\end{equation} represents the \textsc{CNOT imbalance}, i.e., the normalized negative difference between the number of CNOTs in the two partitions. The second term:
\begin{equation}
    \displaystyle \mathcal{L}_2 = -\,\frac{|\Delta_{Cells}|}{N_{q}\,N_{MC}}\,\,\,\,\in\,[-1,0],
\end{equation} represents the \textsc{cell imbalance}, i.e.,the normalized negative difference between the number of maze cells assigned to each of the two partitions, and the third term:
\begin{equation}
    \displaystyle \mathcal{L}_3 = -\,\frac{|l_{path} - N_{MC}|}{N_{q}-1}\,\,\,\,\in\,[-1,0],
\end{equation} computes the normalized path length of the agent through the maze.

In Figure~\ref{fig:maze metrics}, we show the heatmaps of the scores defined in Eq.~\eqref{eq:l_score_tot}, and obtained by selecting the best-performing agent among all the candidates.
The scores are evaluated by varying the relevant discrete parameters of the maze, i.e., the number of qubits (which determines the number of rows in the maze) and the depth of the CNOT circuit (which determines the number of columns in each maze).
In this work, the weights $(\alpha, \beta, \gamma)$ are fixed at $(1,1,2)$.
This choice gives equal weight to the two balancing criteria, namely CNOT imbalance and cell imbalance, while assigning greater weight to the path length penalty.
In this way, the selected cut is encouraged to traverse the maze from left to right with limited vertical deviations.
This is important because a predominantly vertical path would not provide a significant circuit partition, even if it locally avoided the maze walls.
Alternative possible settings are provided in Appendix \ref{appendix:gamma}. 
Each cell in the heatmaps refers to the average value obtained from $1000$ simulations. Each simulation corresponds to the execution of all upward and downward agents for each cell in the first column. Only the cut that maximizes $\mathcal{L}_{tot}$ is taken into account and used as a contribution to the average value. 

Let us consider the length of the shortest path as shown in Figure \ref{fig:maze metrics} (c). 
This metric shows a transition between two distinct phases, represented by two black regions separated by a clearly visible purple central band. 
The first phase, which is shown in Figure \ref{fig:maze metrics} (d), corresponds to the black region on the left (red square) in Figure \ref{fig:maze metrics} (c). 
In this regime, the optimal cut is the one that crosses the maze in the center from left to right without intersecting any walls. 
As a result, the cut is strictly horizontal and produces a balanced partition with respect to both the CNOT imbalance and the difference in the number of maze cells between the two partitions. 
In this case the whole quantum circuit can be represented simply as the tensor product of the two partitions.
The circuits belonging to this region are clearly partitionable according to our method.
Focusing on the right region in Figure \ref{fig:maze metrics} (c) (violet square), which corresponds to the path shown in Figure \ref{fig:maze metrics} (f), the optimal cut is again horizontal. 
However, in this case, the agent proceeds along the boundary of the maze instead of producing a meaningful separation of the maze into two parts.
Indeed, in this regime the agent is almost unable to cut the circuit horizontally far from the upper and lower boundaries due to the presence of dense clusters of walls.
As a result, the cut fails to induce a proper partition of the maze. 
One partition has a negligible area, while the other one contains almost all the cells of the maze and all the CNOT gates of the circuit.
We can therefore consider the circuits analyzed in this regime as belonging to the class of non-partitionable circuits.
In the central region we observe a transitional behavior (green square), which corresponds to the path shown in \ref{fig:maze metrics} (e), in which the length of the best cut tends to increase first and then decreases again. 
Hence, during its traversal of the maze, the best agent encounters several walls that prevent it from moving exclusively horizontally. 
For this reason, in order to reach the destination column, the agent follows a path that alternates vertical and horizontal steps, creating a sequence of moves upwards and to the right or downwards and to the right.
Although not well balanced in the first two metrics evaluated, the partition obtained with this type of cut is still a valid partition in a regime where the SA is no longer able to completely cluster the maze walls.

The phase transition emerges also in the case of $\mathcal{L}_1$ -- see Figure \ref{fig:maze metrics} (a) -- and $\mathcal{L}_2$ -- see Figure \ref{fig:maze metrics} (b) --, but in these cases the behaviour of the metric appears to be smoother than in the case of $\mathcal{L}_3$, with two clear distinct regions corresponding to the different phases. Clearly, the average values of $\mathcal{L}_1$ and $\mathcal{L}_2$ in the second phase are different than the ones in the first phase, which further highlights the radically different nature of these two regimes.

To confirm our results, we also compare them with the commutator between the two circuit partitions obtained with our method.
In Appendix \ref{app:commutator} we study the behaviour of the quantum mechanical commutator of the two partitions, and in Appendix \ref{app:mass transport} we draw a connection between the commutator and optimal transport theory \cite{Rachev1998}, using the so-called observable mass balance.

\section{Conclusions and future perspectives}
\label{section:conclusions}

We introduced a geometric framework for quantum circuit partitioning in which the circuit is mapped onto a maze, with CNOT gates acting as walls.
Using this representation, the partitioning problem for circuit optimization is reformulated as finding a left-to-right path through a maze, without cutting any CNOT gate.
Such reformulation allows us to distinguish, directly and operationally, circuits that admit a suitable partition from those not admitting it.
Above all, our results show that this distinction behaves like the separation between two phases in a percolation dynamics, which emerges after optimizing the qubit ordering via simulated annealing and subsequently performing a circuit cut using deterministic maze-traversal agents.
We identify three regimes, namely two distinct phases and an intermediate regime, which depend on the scale between the CNOT count and the number of qubits in the circuit.

In the first phase, the number of CNOT gates is lower than the number of qubits, and SA rearranges the interaction pattern into two clearly separated CNOT clusters.
Therefore, each quantum circuit can be divided into two subcircuits, and the best path that passes through the center of the maze remains horizontal.
By representing the circuit as an interaction graph, the annealer reshapes it into a loosely coupled structure, with lower conductance, greater community formation, and less mixing of the information flow.

In the second phase these characteristics disappear and the interaction graph becomes strongly connected, preventing any significant separation of the CNOT blocks. 
Thus, the optimal path is pushed towards the boundary of the maze, generating a trivial cut. In this regime, no meaningful partitioning is possible.

The region between these two phases is naturally interpreted as a transition regime connecting a partitionable phase to a non-partitionable one.
In this intermediate phase we observe that there is still a clear distinction between the two clusters of CNOTs, although the cut path itself behaves differently.
The latter consists of both horizontal and vertical steps, causing an increase in path length and making it more difficult to obtain balanced partitions.
This transition regime corresponds to a number of CNOT gates approximately equal to the number of qubits. 
The proportion between these two quantities allows us to establish a criterion to determine when a circuit cut delivers distinct subcircuits that can be optimized independently.
From a graph-theoretical point of view, the overall transition is consistent with a percolation on an Erd\H{o}s--R\'enyi-type interaction graph with a average degree $\langle k \rangle \sim 1$.

Overall, our results provide a principled criterion for determining when partitioning is feasible in the context of quantum circuit optimization, thereby laying the groundwork for further algorithmic development.

\section*{Code availability}
The code and data are available upon request.

\begin{acknowledgments}
We are grateful to Stefan Kesselheim,  Matteo Robbiati, Luca Nigro, Jos\'e Jesus and Dimitrios Georgiadis for the stimulating discussions. This work was supported by AIDAS, The European Joint Virtual Lab, an initiative of Forschungszentrum J\"ulich (FZJ) and the French Alternative Energies and Atomic Energy Commission (CEA), by the German Federal Ministry of Education and Research (BMBF), project QSolid, Grant No.~13N16149, by the German Research Foundation (DFG) under Germany’s Excellence Strategy – Cluster of Excellence Matter and Light for Quantum Computing (ML4Q2) EXC 2004/2 – 390534769 and by the J\"ulich Supercomputing Center (JSC). We acknowledge funding from the Horizon Europe program (HORIZON-CL4-2021-DIGITAL-EMERGING-02-10) via the project 101080085 (QCFD) and by HORIZON-CL4-2022-QUANTUM-01-SGA Project
under Grant 101113946 OpenSuperQPlus100.  Simulations were realized in \textsc{python} using the libraries \textsc{qiskit} \cite{javadi2024quantum}, \textsc{networkx} \cite{hagberg2007exploring}, \textsc{scipy} \cite{gommers2024scipy}, \textsc{qibo} \cite{qibo_paper}.
P. Z. and E. P. are grateful to NTT Data for having co-funded the D.M. 117 PNRR PhD grant.

\end{acknowledgments}

\section*{Contributions}
P. Z. conceived the research, planned the work at all stages and developed the main code of the simulation.
M. G. developed the graph-theoretical framework and the related simulations.
F. P. developed the theoretical framework for SA and the percolation analysis of the maze.
D. A. contributed to the overall development of the project.
F. P., F. A. C. L., F. M. and E. P. supervised the work.
All the authors wrote and revised the manuscript.

\bibliographystyle{apsrev4-2-author-truncate}
\bibliography{bibliography}  

@BOOK{Bullo2018,
  title     = "Lectures on Network Systems",
  author    = "Bullo, Francesco",
  publisher = "Createspace Independent Publishing Platform",
  month     =  mar,
  year      =  2018,
  address   = "North Charleston, SC",
  url = "https://fbullo.github.io/lns/"
}

@book{bishop2006pattern,
  title={Pattern Recognition and Machine Learning},
  author={Bishop, C.M.},
  isbn={9780387310732},
  year={2006},
  publisher={Springer}
}

@book{Rachev1998,
  title     = "Mass Transportation Problems",
  author    = "Rachev, Svetlozar T and Ruschendorf, Ludger",
  publisher = "Springer",
  series    = "Probability and Its Applications",
  edition   =  1998,
  month     =  mar,
  year      =  1998,
  address   = "New York, NY",
}

@inbook{Espanol2004,
author = {Español, Pep},
year = {2004},
month = {02},
pages = {2256-2256},
title = {Statistical Mechanics of Coarse-Graining},
volume = {640},
isbn = {978-3-540-20916-4},
journal = {Lect. Notes Phys.},
doi = {10.1007/978-3-540-39895-0_3}
}

@book{CastiglioneCoarseGraining, place={Cambridge}, title={Chaos and Coarse Graining in Statistical Mechanics}, publisher={Cambridge University Press}, author={Castiglione, Patrizia and Falcioni, Massimo and Lesne, Annick and Vulpiani, Angelo}, year={2008}}

@article{Thaler_2022,
   title={Deep coarse-grained potentials via relative entropy minimization},
   volume={157},
   ISSN={1089-7690},
   url={http://dx.doi.org/10.1063/5.0124538},
   doi={10.1063/5.0124538},
   number={24},
   journal={The Journal of Chemical Physics},
   publisher={AIP Publishing},
   author={Thaler, Stephan and Stupp, Maximilian and Zavadlav, Julija},
   year={2022},
   month=dec }

@book{Butler2007, place={Cambridge}, series={Cambridge Series in Statistical and Probabilistic Mathematics}, title={Saddlepoint Approximations with Applications}, publisher={Cambridge University Press}, author={Butler, Ronald W.}, year={2007}, collection={Cambridge Series in Statistical and Probabilistic Mathematics}}

@misc{majdalani_laplace_method,
  author = {Majdalani, J.},
  title = {Lecture on the Laplace Method},
  year = {},
  url = {http://majdalani.eng.auburn.edu/courses/06_perturbations_2/Lecture_on_the_Laplace_Method.pdf},
  note = {Accessed: 2026-03-30}
}

@article{kirkpatrick1983optimization,
  title={Optimization by simulated annealing},
  author={Kirkpatrick, S and Gelatt Jr, C D and Vecchi, M P},
  journal={Science},
  volume={220},
  number={4598},
  pages={671--680},
  year={1983},
  publisher={AAAS},
  url={https://www.science.org/doi/10.1126/science.220.4598.671}
}

@article{Kissinger_2020,
   title={PyZX: Large Scale Automated Diagrammatic Reasoning},
   volume={318},
   ISSN={2075-2180},
   url={http://dx.doi.org/10.4204/EPTCS.318.14},
   doi={10.4204/eptcs.318.14},
   journal={Electronic Proceedings in Theoretical Computer Science},
   publisher={Open Publishing Association},
   author={Kissinger, Aleks and van de Wetering, John},
   year={2020},
   month=may, pages={229–241} }

@misc{nemirovsky2025efficientcompilationquantumcircuits,
      title={Efficient compilation of quantum circuits using multi-qubit gates}, 
      author={Jonathan Nemirovsky and Maya Chuchem and Yotam Shapira},
      year={2025},
      eprint={2501.17246},
      archivePrefix={arXiv},
      primaryClass={quant-ph},
}

@misc{zulehner2018efficientmethodologymappingquantum,
      title={An Efficient Methodology for Mapping Quantum Circuits to the IBM QX Architectures}, 
      author={Alwin Zulehner and Alexandru Paler and Robert Wille},
      year={2018},
      eprint={1712.04722},
      archivePrefix={arXiv},
      primaryClass={quant-ph},
}

@article{Ender2023,
   title={Parity Quantum Optimization: Compiler},
   volume={7},
   ISSN={2521-327X},
   url={http://dx.doi.org/10.22331/q-2023-03-17-950},
   doi={10.22331/q-2023-03-17-950},
   journal={Quantum},
   publisher={Verein zur Forderung des Open Access Publizierens in den Quantenwissenschaften},
   author={Ender, Kilian and ter Hoeven, Roeland and Niehoff, Benjamin E. and Drieb-Schön, Maike and Lechner, Wolfgang},
   year={2023},
   month=mar, pages={950} }

@misc{BQSKIT,
    title = {Berkeley Quantum Synthesis Toolkit (BQSKit) v1},
    author = {Younis, Ed and Iancu, Costin C. and Lavrijsen, Wim and Davis, Marc and Smith, Ethan},
    abstractNote = {The Berkeley Quantum Synthesis Toolkit (BQSKit) is an optimizing quantum compiler and research vehicle that combines ideas from several projects at LBNL into one easily accessible and quickly extensible software package. The ideas in the QFAST, QSearch, LEAP, and QFactor software tools (all licensed through ipo.lbl.gov) all build upon one another. By combining these into one package, we create symbiotic interactions between the tools. This means better results, better throughput, less to maintain, and greater surface area to the public. Additionally, the BQSKit tool will create a research platform for future work here at LBNL.},
    doi = {10.11578/dc.20210603.2},
    url = {https://doi.org/10.11578/dc.20210603.2},
    howpublished = {[Computer Software] \url{https://doi.org/10.11578/dc.20210603.2}},
    year = {2021},
    month = {apr}
}

@article{moro2021quantum,
  author    = {Moro, Lorenzo and Paris, Matteo G. A. and Restelli, Marcello and Prati, Enrico},
  title     = {Quantum Compiling by Deep Reinforcement Learning},
  journal   = {Communications Physics},
  volume    = {4},
  number    = {1},
  pages     = {178},
  year      = {2021},
  doi       = {10.1038/s42005-021-00684-3},
  publisher = {Nature Publishing Group}
}

@misc{kremer2024practical,
  title={Practical and efficient quantum circuit synthesis and transpiling with Reinforcement Learning}, 
  author={David Kremer and Victor Villar and Hanhee Paik and Ivan Duran and Ismael Faro and Juan Cruz-Benito},
  year={2025},
  eprint={2405.13196},
  archivePrefix={arXiv},
  primaryClass={quant-ph},
}

@article{corli2026measurement,
  title={Measurement-based quantum compiling via gauge invariance},
  author={Corli, Sebastiano and Prati, Enrico},
  journal={Physical Review Applied},
  volume={25},
  number={4},
  pages={044068},
  year={2026},
  publisher={APS}
}

@inproceedings{corli2025gauge,
  title={Gauge freedom in measurement based quantum compiling},
  author={Corli, Sebastiano and Prati, Enrico},
  booktitle={Journal of Physics: Conference Series},
  volume={3017},
  number={1},
  pages={012043},
  year={2025},
  organization={IOP Publishing}
}

@article{mele2024introduction,
   title={Introduction to Haar Measure Tools in Quantum Information: A Beginner's Tutorial},
   volume={8},
   ISSN={2521-327X},
   url={http://dx.doi.org/10.22331/q-2024-05-08-1340},
   DOI={10.22331/q-2024-05-08-1340},
   journal={Quantum},
   publisher={Verein zur Forderung des Open Access Publizierens in den Quantenwissenschaften},
   author={Mele, Antonio Anna},
   year={2024},
   month=may, pages={1340} }

@article{brennen2003observable,
  title={An observable measure of entanglement for pure states of multi-qubit systems},
  author={Brennen, Gavin K},
  journal={arXiv preprint},
  eprint={arXiv:0305094},
  year={2003}
}

@article{meyer2001global,
  title={Global entanglement in multiparticle systems},
  author={Meyer, David A and Wallach, Nolan R},
  archivePrefix={arXiv},
  eprint={arXiv:0108104},
  journal={arXiv preprint},
  year={2001}
}

@article{Nahum2021,
   title={Measurement and Entanglement Phase Transitions in All-To-All Quantum Circuits, on Quantum Trees, and in Landau-Ginsburg Theory},
   volume={2},
   ISSN={2691-3399},
   number={1},
   journal={PRX Quantum},
   publisher={American Physical Society (APS)},
   author={Nahum, Adam and Roy, Sthitadhi and Skinner, Brian and Ruhman, Jonathan},
   year={2021},
   month=mar,
   url={http://dx.doi.org/10.1103/PRXQuantum.2.010352} }

@article{akahoshi2024partially,
  title={Partially fault-tolerant quantum computing architecture with error-corrected Clifford gates and space-time efficient analog rotations},
  author={Akahoshi, Yutaro and Maruyama, Kazunori and Oshima, Hirotaka and Sato, Shintaro and Fujii, Keisuke},
  journal={PRX quantum},
  volume={5},
  number={1},
  pages={010337},
  year={2024},
  publisher={APS}
}

@article{Barabasi1999,
   title={Emergence of Scaling in Random Networks},
   volume={286},
   ISSN={1095-9203},
   url={http://dx.doi.org/10.1126/science.286.5439.509},
   doi={10.1126/science.286.5439.509},
   number={5439},
   journal={Science},
   publisher={American Association for the Advancement of Science (AAAS)},
   author={Barabási, Albert-László and Albert, Réka},
   year={1999},
   month=oct, pages={509–512} }

@article{Bollobas1980,
title = {A Probabilistic Proof of an Asymptotic Formula for the Number of Labelled Regular Graphs},
journal = {European Journal of Combinatorics},
volume = {1},
number = {4},
pages = {311-316},
year = {1980},
issn = {0195-6698},
doi = {https://doi.org/10.1016/S0195-6698(80)80030-8},
url = {https://www.sciencedirect.com/science/article/pii/S0195669880800308},
author = {Béla Bollobás}
}

@article{erdos59a,
  author = {Erd\"{o}s, P. and R\'{e}nyi, A.},
  journal = {Publicationes Mathematicae Debrecen},
  pages = 290,
  title = {On Random Graphs I},
  volume = 6,
  year = 1959
}

@article{erdHos1960evolution,
  title={On the evolution of random graphs},
  author={Erd{\H{o}}s, Paul and R{\'e}nyi, Alfr{\'e}d and others},
  journal={Publications of the},
  year={1960}
}

@article{djordjevic1982site,
  title={Site percolation threshold for honeycomb and square lattices},
  author={Djordjevic, Zorica V and Stanley, H Eugene and Margolina, Alla},
  journal={Journal of Physics A: Mathematical and General},
  volume={15},
  number={8},
  pages={L405--L412},
  year={1982}
}

@book{barabasi2016network,
  address = {Cambridge},
  author = {Barabási, Albert-László and Pósfai, Márton},
  description = {Network Science: Albert-László Barabási: 9781107076266: Amazon.com: Books},
  isbn = {9781107076266 1107076269},
  publisher = {Cambridge University Press},
  refid = {958874494},
  title = {Network science},
  url = {http://barabasi.com/networksciencebook/},
  year = 2016
}

@article{degroot1974reaching,
 ISSN = {01621459, 1537274X},
 URL = {http://www.jstor.org/stable/2285509},
 author = {Morris H. DeGroot},
 journal = {Journal of the American Statistical Association},
 number = {345},
 pages = {118--121},
 publisher = {[American Statistical Association, Taylor & Francis, Ltd.]},
 title = {Reaching a Consensus},
 urldate = {2026-04-01},
 volume = {69},
 year = {1974}
}

@article{barthelemy2004betweenness,
  title={Betweenness centrality in large complex networks},
  author={Barthelemy, Marc},
  journal={The European physical journal B},
  volume={38},
  number={2},
  pages={163--168},
  year={2004},
  publisher={Springer}
}

@misc{banfi2025transpiling,
  title={Transpiling quantum circuits by a transformers-based algorithm}, 
  author={Michele Banfi and Paolo Zentilini and Sebastiano Corli and Enrico Prati},
  year={2026},
  eprint={2512.09834},
  archivePrefix={arXiv},
  primaryClass={quant-ph},
}

@misc{tejedor2025distributed,
  title={Distributed Quantum Circuit Cutting for Hybrid Quantum-Classical High-Performance Computing}, 
  author={Mar Tejedor and Berta Casas and Javier Conejero and Alba Cervera-Lierta and Rosa M. Badia},
  year={2025},
  eprint={2505.01184},
  archivePrefix={arXiv},
  primaryClass={cs.DC},
}

@inproceedings{xu2025optimizing,
  title={Optimizing quantum circuits, fast and slow},
  author={Xu, Amanda and Molavi, Abtin and Tannu, Swamit and Albarghouthi, Aws},
  booktitle={Proceedings of the 30th ACM International Conference on Architectural Support for Programming Languages and Operating Systems, Volume 1},
  pages={777--793},
  year={2025}
}

@article{daei2020optimized,
   title={Optimized Quantum Circuit Partitioning},
   volume={59},
   ISSN={1572-9575},
   url={http://dx.doi.org/10.1007/s10773-020-04633-8},
   number={12},
   journal={International Journal of Theoretical Physics},
   publisher={Springer Science and Business Media LLC},
   author={Daei, Omid and Navi, Keivan and Zomorodi-Moghadam, Mariam},
   year={2020},
   month=nov, pages={3804–3820} }

@misc{wu2020qgo,
  title={QGo: Scalable Quantum Circuit Optimization Using Automated Synthesis}, 
  author={Xin-Chuan Wu and Marc Grau Davis and Frederic T. Chong and Costin Iancu},
  year={2022},
  archivePrefix={arXiv},
  eprint={2012.09835},
  primaryClass={quant-ph},
}

@misc{tamura2026decompositionmultiqubitgatescircuit,
      title={Decomposition of Multi-Qubit Gates for Circuit Cutting}, 
      author={Ryota Tamura and Tomoya Kashimata and Yohei Hamakawa and Kosuke Tatsumura and Hiroshi Imai},
      year={2026},
      eprint={2603.26278},
      archivePrefix={arXiv},
      primaryClass={quant-ph}, 
}

@article{burt2026multilevel,
  title={A multilevel framework for partitioning quantum circuits},
  author={Burt, Felix and Chen, Kuan-Cheng and Leung, Kin K},
  journal={Quantum},
  volume={10},
  pages={1984},
  year={2026},
  publisher={Verein zur F{\"o}rderung des Open Access Publizierens in den Quantenwissenschaften}
}

@misc{javadi2024quantum,
 title={Quantum computing with Qiskit}, 
      author={Ali Javadi-Abhari and Matthew Treinish and Kevin Krsulich and Christopher J. Wood and Jake Lishman and Julien Gacon and Simon Martiel and Paul D. Nation and Lev S. Bishop and Andrew W. Cross and Blake R. Johnson and Jay M. Gambetta},
      year={2024},
      eprint={2405.08810},
      archivePrefix={arXiv},
      primaryClass={quant-ph},
}

@article{qibo_paper,
    doi       = {10.1088/2058-9565/ac39f5},
    url       = {https://doi.org/10.1088/2058-9565/ac39f5},
    year      = 2021,
    month     = {dec},
    publisher = {{IOP} Publishing},
    volume    = {7},
    number    = {1},
    pages     = {015018},
    author    = {Stavros Efthymiou and
                 Sergi Ramos-Calderer and
                 Carlos Bravo-Prieto and
                 Adri{\'{a}}n P{\'{e}}rez-Salinas and
                 Diego Garc{\'{\i}}a-Mart{\'{\i}}n and
                 Artur Garcia-Saez and
                 Jos{\'{e}} Ignacio Latorre and
                 Stefano Carrazza},
    title     = {Qibo: a framework for quantum simulation with hardware acceleration},
    journal   = {Quantum Science and Technology},
}

@misc{gommers2024scipy,
  title={scipy/scipy: SciPy 1.15. 0},
  author={Gommers, Ralf and Virtanen, Pauli and Haberland, Matt and Burovski, Evgeni and Reddy, Tyler and Weckesser, Warren and Oliphant, Travis E and Cournapeau, David and Nelson, Andrew and Roy, Pamphile and others},
  journal={Zenodo},
  url={https://zenodo.org/records/14552608},
  year={2024}
}

@techreport{hagberg2007exploring,
  title={Exploring network structure, dynamics, and function using NetworkX},
  author={Hagberg, Aric and Swart, Pieter J and Schult, Daniel A},
  year={2007},
  institution={Los Alamos National Laboratory (LANL)}
}

@article{storn1997differential,
author  = {Storn, Rainer and Price, Kenneth},
  title   = {Differential Evolution -- A Simple and Efficient Heuristic for Global Optimization over Continuous Spaces},
  journal = {Journal of Global Optimization},
  volume  = {11},
  number  = {4},
  pages   = {341--359},
  year    = {1997},
  doi     = {10.1023/A:1008202821328}
}

@article{Jian2020,
  title = {Measurement-induced criticality in random quantum circuits},
  author = {Jian, Chao-Ming and You, Yi-Zhuang and Vasseur, Romain and Ludwig, Andreas W. W.},
  journal = {Phys. Rev. B},
  volume = {101},
  issue = {10},
  pages = {104302},
  numpages = {11},
  year = {2020},
  month = {Mar},
  publisher = {American Physical Society},
  doi = {10.1103/PhysRevB.101.104302},
  url = {https://link.aps.org/doi/10.1103/PhysRevB.101.104302}
}

@article{Buznach2025,
  title = {Unveiling a Hidden Percolation Transition in Monitored Clifford Circuits: Inroads from ZX Calculus},
  author = {Buznach Ahituv, Einat and Chowdhury, Debanjan and Ruhman, Jonathan},
  journal = {Phys. Rev. Lett.},
  volume = {135},
  issue = {5},
  pages = {050402},
  numpages = {7},
  year = {2025},
  month = {Jul},
  publisher = {American Physical Society},
  doi = {10.1103/mgy3-kr71},
  url = {https://link.aps.org/doi/10.1103/mgy3-kr71}
}

@article{zabalo2020critical,
  title={Critical properties of the measurement-induced transition in random quantum circuits},
  author={Zabalo, Aidan and Gullans, Michael J and Wilson, Justin H and Gopalakrishnan, Sarang and Huse, David A and Pixley, JH},
  journal={Physical Review B},
  volume={101},
  number={6},
  pages={060301},
  year={2020},
  publisher={APS}
}

@article{choi2020quantum,
  title={Quantum error correction in scrambling dynamics and measurement-induced phase transition},
  author={Choi, Soonwon and Bao, Yimu and Qi, Xiao-Liang and Altman, Ehud},
  journal={Physical Review Letters},
  volume={125},
  number={3},
  pages={030505},
  year={2020},
  publisher={APS}
}

@article{li2018quantum,
  title={Quantum Zeno effect and the many-body entanglement transition},
  author={Li, Yaodong and Chen, Xiao and Fisher, Matthew PA},
  journal={Physical Review B},
  volume={98},
  number={20},
  pages={205136},
  year={2018},
  publisher={APS}
}

@article{lavasani2021measurement,
  title={Measurement-induced topological entanglement transitions in symmetric random quantum circuits},
  author={Lavasani, Ali and Alavirad, Yahya and Barkeshli, Maissam},
  journal={Nature Physics},
  volume={17},
  number={3},
  pages={342--347},
  year={2021},
  publisher={Nature Publishing Group UK London}
}

@article{jonay2021triunitary,
  title={Triunitary quantum circuits},
  author={Jonay, Cheryne and Khemani, Vedika and Ippoliti, Matteo},
  journal={Physical Review Research},
  volume={3},
  number={4},
  pages={043046},
  year={2021},
  publisher={APS}
}

@article{ruiz2025quantum,
  title={Quantum circuit optimization with alphatensor},
  author={Ruiz, Francisco JR and Laakkonen, Tuomas and Bausch, Johannes and Balog, Matej and Barekatain, Mohammadamin and Heras, Francisco JH and Novikov, Alexander and Fitzpatrick, Nathan and Romera-Paredes, Bernardino and Van De Wetering, John and others},
  journal={Nature Machine Intelligence},
  volume={7},
  number={3},
  pages={374--385},
  year={2025},
  publisher={Nature Publishing Group UK London}
}

@article{nam2018automated,
  title={Automated optimization of large quantum circuits with continuous parameters},
  author={Nam, Yunseong and Ross, Neil J and Su, Yuan and Childs, Andrew M and Maslov, Dmitri},
  journal={npj Quantum Information},
  volume={4},
  number={1},
  pages={23},
  year={2018},
  publisher={Nature Publishing Group UK London}
}

@article{Amy_2014,
   title={Polynomial-Time T-Depth Optimization of Clifford+T Circuits Via Matroid Partitioning},
   volume={33},
   ISSN={1937-4151},
   url={http://dx.doi.org/10.1109/TCAD.2014.2341953},
   doi={10.1109/tcad.2014.2341953},
   number={10},
   journal={IEEE Transactions on Computer-Aided Design of Integrated Circuits and Systems},
   publisher={Institute of Electrical and Electronics Engineers (IEEE)},
   author={Amy, Matthew and Maslov, Dmitri and Mosca, Michele},
   year={2014},
   month=oct, pages={1476–1489} 
}

@article{lanyon2010towards,
  title={Towards quantum chemistry on a quantum computer},
  author={Lanyon, Benjamin P and Whitfield, James D and Gillett, Geoff G and Goggin, Michael E and Almeida, Marcelo P and Kassal, Ivan and Biamonte, Jacob D and Mohseni, Masoud and Powell, Ben J and Barbieri, Marco and others},
  journal={Nature chemistry},
  volume={2},
  number={2},
  pages={106--111},
  year={2010},
  publisher={Nature Publishing Group}
}

@inproceedings{Siraichi2018,
    author = {Siraichi, Marcos Yukio and Santos, Vin\'{\i}cius Fernandes dos and Collange, Caroline and Pereira, Fernando Magno Quintao},
    title = {Qubit allocation},
    year = {2018},
    isbn = {9781450356176},
    publisher = {Association for Computing Machinery},
    address = {New York, NY, USA},
    url = {https://doi.org/10.1145/3168822},
    doi = {10.1145/3168822},
    booktitle = {Proceedings of the 2018 International Symposium on Code Generation and Optimization},
    pages = {113–125},
    numpages = {13},
    location = {Vienna, Austria},
    series = {CGO '18}
    }

@article{Liu2024,
author={Liu, Junyu
and Liu, Minzhao
and Liu, Jin-Peng
and Ye, Ziyu
and Wang, Yunfei
and Alexeev, Yuri
and Eisert, Jens
and Jiang, Liang},
title={Towards provably efficient quantum algorithms for large-scale machine-learning models},
journal={Nature Communications},
year={2024},
month={Jan},
day={10},
volume={15},
number={1},
pages={434},
issn={2041-1723},
doi={10.1038/s41467-023-43957-x},
url={https://doi.org/10.1038/s41467-023-43957-x}
}

@article{Acharya2025,
author={Acharya, Rajeev
and Abanin, Dmitry A.
and Aghababaie-Beni, Laleh
and Aleiner, Igor
and Andersen, Trond I.
and Ansmann, Markus
and Arute, Frank
and Arya, Kunal
and Asfaw, Abraham
and Astrakhantsev, Nikita
and Atalaya, Juan
and Babbush, Ryan
and Bacon, Dave
and Ballard, Brian
and Bardin, Joseph C.
and Bausch, Johannes
and Bengtsson, Andreas
and Bilmes, Alexander
and Blackwell, Sam
and Boixo, Sergio
and Bortoli, Gina
and Bourassa, Alexandre
and Bovaird, Jenna
and Brill, Leon
and Broughton, Michael
and Browne, David A.
and Buchea, Brett
and Buckley, Bob B.
and Buell, David A.
and Burger, Tim
and Burkett, Brian
and Bushnell, Nicholas
and Cabrera, Anthony
and Campero, Juan
and Chang, Hung-Shen
and Chen, Yu
and Chen, Zijun
and Chiaro, Ben
and Chik, Desmond
and Chou, Charina
and Claes, Jahan
and Cleland, Agnetta Y.
and Cogan, Josh
and Collins, Roberto
and Conner, Paul
and Courtney, William
and Crook, Alexander L.
and Curtin, Ben
and Das, Sayan
and Davies, Alex
and De Lorenzo, Laura
and Debroy, Dripto M.
and Demura, Sean
and Devoret, Michel
and Di Paolo, Agustin
and Donohoe, Paul
and Drozdov, Ilya
and Dunsworth, Andrew
and Earle, Clint
and Edlich, Thomas
and Eickbusch, Alec
and Elbag, Aviv Moshe
and Elzouka, Mahmoud
and Erickson, Catherine
and Faoro, Lara
and Farhi, Edward
and Ferreira, Vinicius S.
and Burgos, Leslie Flores
and Forati, Ebrahim
and Fowler, Austin G.
and Foxen, Brooks
and Ganjam, Suhas
and Garcia, Gonzalo
and Gasca, Robert
and Genois, {\'E}lie
and Giang, William
and Gidney, Craig
and Gilboa, Dar
and Gosula, Raja
and Dau, Alejandro Grajales
and Graumann, Dietrich
and Greene, Alex
and Gross, Jonathan A.
and Habegger, Steve
and Hall, John
and Hamilton, Michael C.
and Hansen, Monica
and Harrigan, Matthew P.
and Harrington, Sean D.
and Heras, Francisco J. H.
and Heslin, Stephen
and Heu, Paula
and Higgott, Oscar
and Hill, Gordon
and Hilton, Jeremy
and Holland, George
and Hong, Sabrina
and Huang, Hsin-Yuan
and Huff, Ashley
and Huggins, William J.
and Ioffe, Lev B.
and Isakov, Sergei V.
and Iveland, Justin
and Jeffrey, Evan
and Jiang, Zhang
and Jones, Cody
and Jordan, Stephen
and Joshi, Chaitali
and Juhas, Pavol
and Kafri, Dvir
and Kang, Hui
and Karamlou, Amir H.
and Kechedzhi, Kostyantyn
and Kelly, Julian
and Khaire, Trupti
and Khattar, Tanuj
and Khezri, Mostafa
and Kim, Seon
and Klimov, Paul V.
and Klots, Andrey R.
and Kobrin, Bryce
and Kohli, Pushmeet
and Korotkov, Alexander N.
and Kostritsa, Fedor
and Kothari, Robin
and Kozlovskii, Borislav
and Kreikebaum, John Mark
and Kurilovich, Vladislav D.
and Lacroix, Nathan
and Landhuis, David
and Lange-Dei, Tiano
and Langley, Brandon W.
and Laptev, Pavel
and Lau, Kim-Ming
and Le Guevel, Lo{\"i}ck
and Ledford, Justin
and Lee, Joonho
and Lee, Kenny
and Lensky, Yuri D.
and Leon, Shannon
and Lester, Brian J.
and Li, Wing Yan
and Li, Yin
and Lill, Alexander T.
and Liu, Wayne
and Livingston, William P.
and Locharla, Aditya
and Lucero, Erik
and Lundahl, Daniel
and Lunt, Aaron
and Madhuk, Sid
and Malone, Fionn D.
and Maloney, Ashley
and Mandr{\`a}, Salvatore
and Manyika, James
and Martin, Leigh S.
and Martin, Orion
and Martin, Steven
and Maxfield, Cameron
and McClean, Jarrod R.
and McEwen, Matt
and Meeks, Seneca
and Megrant, Anthony
and Mi, Xiao
and Miao, Kevin C.
and Mieszala, Amanda
and Molavi, Reza
and Molina, Sebastian
and Montazeri, Shirin
and Morvan, Alexis
and Movassagh, Ramis
and Mruczkiewicz, Wojciech
and Naaman, Ofer
and Neeley, Matthew
and Neill, Charles
and Nersisyan, Ani
and Neven, Hartmut
and Newman, Michael
and Ng, Jiun How
and Nguyen, Anthony
and Nguyen, Murray
and Ni, Chia-Hung
and Niu, Murphy Yuezhen
and O'Brien, Thomas E.
and Oliver, William D.
and Opremcak, Alex
and Ottosson, Kristoffer
and Petukhov, Andre
and Pizzuto, Alex
and Platt, John
and Potter, Rebecca
and Pritchard, Orion
and Pryadko, Leonid P.
and Quintana, Chris
and Ramachandran, Ganesh
and Reagor, Matthew J.
and Redding, John
and Rhodes, David M.
and Roberts, Gabrielle
and Rosenberg, Eliott
and Rosenfeld, Emma
and Roushan, Pedram
and Rubin, Nicholas C.
and Saei, Negar
and Sank, Daniel
and Sankaragomathi, Kannan
and Satzinger, Kevin J.
and Schurkus, Henry F.
and Schuster, Christopher
and Senior, Andrew W.
and Shearn, Michael J.
and Shorter, Aaron
and Shutty, Noah
and Shvarts, Vladimir
and Singh, Shraddha
and Sivak, Volodymyr
and Skruzny, Jindra
and Small, Spencer
and Smelyanskiy, Vadim
and Smith, W. Clarke
and Somma, Rolando D.
and Springer, Sofia
and Sterling, George
and Strain, Doug
and Suchard, Jordan
and Szasz, Aaron
and Sztein, Alex
and Thor, Douglas
and Torres, Alfredo
and Torunbalci, M. Mert
and Vaishnav, Abeer
and Vargas, Justin
and Vdovichev, Sergey
and Vidal, Guifre
and Villalonga, Benjamin
and Heidweiller, Catherine Vollgraff
and Waltman, Steven
and Wang, Shannon X.
and Ware, Brayden
and Weber, Kate
and Weidel, Travis
and White, Theodore
and Wong, Kristi
and Woo, Bryan W. K.
and Xing, Cheng
and Yao, Z. Jamie
and Yeh, Ping
and Ying, Bicheng
and Yoo, Juhwan
and Yosri, Noureldin
and Young, Grayson
and Zalcman, Adam
and Zhang, Yaxing
and Zhu, Ningfeng
and Zobrist, Nicholas
and AI, Google Quantum
and {Collaborators}},
title={Quantum error correction below the surface code threshold},
journal={Nature},
year={2025},
month={Feb},
day={01},
volume={638},
number={8052},
pages={920-926},
issn={1476-4687},
doi={10.1038/s41586-024-08449-y},
url={https://doi.org/10.1038/s41586-024-08449-y}
}

@article{Sauer2021,
doi = {10.1088/1751-8121/ac3469},
url = {https://doi.org/10.1088/1751-8121/ac3469},
year = {2021},
month = {nov},
publisher = {IOP Publishing},
volume = {54},
number = {49},
pages = {495302},
author = {Sauer, A and Bernád, J Z and Moreno, H J and Alber, G},
title = {Entanglement in bipartite quantum systems: Euclidean volume ratios and detectability by Bell inequalities},
journal = {Journal of Physics A: Mathematical and Theoretical}
}

@techreport{Ruane2025QuantumIndex,
  author       = {Ruane, John and Kiesow, Elizabeth and Galatsanos, John and Dukatz, Chris and Blomquist, Erik and Shukla, Pranav},
  title        = {The Quantum Index Report 2025},
  institution  = {MIT Initiative on the Digital Economy, Massachusetts Institute of Technology},
  address      = {Cambridge, MA},
  month        = may,
  year         = {2025}
}

@misc{wang2025circuitdesignstarshapedspinqubit,
      title={Circuit Design for a Star-shaped Spin-Qubit Processor via Algebraic Decomposition and Optimal Control}, 
      author={Yaqing X. Wang and Tommaso Calarco and Felix Motzoi and Matthias M. Müller},
      year={2025},
      eprint={2506.16900},
      archivePrefix={arXiv},
      primaryClass={quant-ph},
}

@misc{guatto2025,
      title={Real-time adaptive quantum error correction by model-free multi-agent learning}, 
      author={Manuel Guatto and Francesco Preti and Michael Schilling and Tommaso Calarco and Francisco Andrés Cárdenas-López and Felix Motzoi},
      year={2025},
      eprint={2509.03974},
      archivePrefix={arXiv},
      primaryClass={quant-ph}
}

@article{Preti2024hybriddiscrete,
  doi = {10.22331/q-2024-05-14-1343},
  url = {https://doi.org/10.22331/q-2024-05-14-1343},
  title = {Hybrid discrete-continuous compilation of trapped-ion quantum circuits with deep reinforcement learning},
  author = {Preti, Francesco and Schilling, Michael and Jerbi, Sofiene and Trenkwalder, Lea M. and Nautrup, Hendrik Poulsen and Motzoi, Felix and Briegel, Hans J.},
  journal = {{Quantum}},
  issn = {2521-327X},
  publisher = {{Verein zur F{\"{o}}rderung des Open Access Publizierens in den Quantenwissenschaften}},
  volume = {8},
  pages = {1343},
  month = may,
  year = {2024}
}

@misc{preti2026gradients,
      title={Gradients, parallelism, and variance of quantum estimates}, 
      author={Francesco Preti and Michael Schilling and József Zsolt Bernád and Tommaso Calarco and Francisco Cárdenas-López and Felix Motzoi},
      year={2026},
      eprint={2509.11214},
      archivePrefix={arXiv},
      primaryClass={quant-ph},
}

@article{Sivak_2023,
   title={Real-time quantum error correction beyond break-even},
   volume={616},
   ISSN={1476-4687},
   url={http://dx.doi.org/10.1038/s41586-023-05782-6},
   doi={10.1038/s41586-023-05782-6},
   number={7955},
   journal={Nature},
   publisher={Springer Science and Business Media LLC},
   author={Sivak, V. V. and Eickbusch, A. and Royer, B. and Singh, S. and Tsioutsios, I. and Ganjam, S. and Miano, A. and Brock, B. L. and Ding, A. Z. and Frunzio, L. and Girvin, S. M. and Schoelkopf, R. J. and Devoret, M. H.},
   year={2023},
   month=mar, pages={50–55} }

@misc{aharonov1999faulttolerantquantumcomputationconstant,
      title={Fault-Tolerant Quantum Computation With Constant Error Rate}, 
      author={Dorit Aharonov and Michael Ben-Or},
      year={1999},
      eprint={quant-ph/9906129},
      archivePrefix={arXiv},
      primaryClass={quant-ph},
}

@article{Li2018quantumzeno,
  title = {Quantum Zeno effect and the many-body entanglement transition},
  author = {Li, Yaodong and Chen, Xiao and Fisher, Matthew P. A.},
  journal = {Phys. Rev. B},
  volume = {98},
  issue = {20},
  pages = {205136},
  numpages = {9},
  year = {2018},
  month = {Nov},
  publisher = {American Physical Society},
  doi = {10.1103/PhysRevB.98.205136},
  url = {https://link.aps.org/doi/10.1103/PhysRevB.98.205136}
}

@article{Skinner2019,
  title = {Measurement-Induced Phase Transitions in the Dynamics of Entanglement},
  author = {Skinner, Brian and Ruhman, Jonathan and Nahum, Adam},
  journal = {Phys. Rev. X},
  volume = {9},
  issue = {3},
  pages = {031009},
  numpages = {21},
  year = {2019},
  month = {Jul},
  publisher = {American Physical Society},
  doi = {10.1103/PhysRevX.9.031009},
  url = {https://link.aps.org/doi/10.1103/PhysRevX.9.031009}
}

@article{Lavasani2021,
   title={Topological Order and Criticality in 
 Monitored Random Quantum Circuits},
   volume={127},
   ISSN={1079-7114},
   number={23},
   journal={Physical Review Letters},
   publisher={American Physical Society (APS)},
   author={Lavasani, Ali and Alavirad, Yahya and Barkeshli, Maissam},
   year={2021},
   month=dec,
   url={https://doi.org/10.1103/PhysRevLett.127.235701}}

@article{KITAEV20032,
title = {Fault-tolerant quantum computation by anyons},
journal = {Annals of Physics},
volume = {303},
number = {1},
pages = {2-30},
year = {2003},
issn = {0003-4916},
doi = {https://doi.org/10.1016/S0003-4916(02)00018-0},
url = {https://www.sciencedirect.com/science/article/pii/S0003491602000180},
author = {A.Yu. Kitaev}
}

@misc{beaudoin2024altgraphredesigningquantumcircuits,
      title={AltGraph: Redesigning Quantum Circuits Using Generative Graph Models for Efficient Optimization}, 
      author={Collin Beaudoin and Koustubh Phalak and Swaroop Ghosh},
      year={2024},
      eprint={2403.12979},
      archivePrefix={arXiv},
      primaryClass={quant-ph},
}

@article{Mazzarella_2015,
   title={Consensus for Quantum Networks: Symmetry From Gossip Interactions},
   volume={60},
   ISSN={1558-2523},
   url={http://dx.doi.org/10.1109/TAC.2014.2336351},
   doi={10.1109/tac.2014.2336351},
   number={1},
   journal={IEEE Transactions on Automatic Control},
   publisher={Institute of Electrical and Electronics Engineers (IEEE)},
   author={Mazzarella, Luca and Sarlette, Alain and Ticozzi, Francesco},
   year={2015},
   month=jan, pages={158–172} }

@article{Ticozzi_2016,
   title={Symmetrizing quantum dynamics beyond gossip-type algorithms},
   volume={74},
   ISSN={0005-1098},
   url={http://dx.doi.org/10.1016/j.automatica.2016.06.019},
   doi={10.1016/j.automatica.2016.06.019},
   journal={Automatica},
   publisher={Elsevier BV},
   author={Ticozzi, Francesco},
   year={2016},
   month=dec, pages={38–46} }

@article{huang2024network_entanglement,
  author       = {Huang, Yiming and Wang, Hao and Ren, Xiao-Long and Lü, Linyuan},
  title        = {Identifying Key Players in Complex Networks via Network Entanglement},
  journal      = {Communications Physics},
  year         = {2024},
  volume       = {7},
  number       = {1},
  pages        = {19},
  doi          = {10.1038/s42005-023-01483-8},
  publisher    = {Nature Publishing Group}
}

@article{sunderhauf2018localization,
  title={Localization with random time-periodic quantum circuits},
  author={S{\"u}nderhauf, Christoph and P{\'e}rez-Garc{\'\i}a, David and Huse, David A and Schuch, Norbert and Cirac, J Ignacio},
  journal={Physical Review B},
  volume={98},
  number={13},
  pages={134204},
  year={2018},
  publisher={APS}
}

\appendix

\section{Preliminaries}

\subsection{Abbreviations and notation}

For clarity, the main abbreviations and mathematical symbols adopted in this work are summarized below:

\begin{table}[ht]
\centering
\label{tab:notation}
\begin{tabular}{ll}
\hline
\textbf{Notation} \\
\hline
$\mathcal{W}$ & wire representation \\
$\mathcal{M}$ & maze representation \\
$N_q$ & number of qubits \\
$N_c$ & number of CNOT gates \\
$N_{\text{MC}}$ & number of  columns in $\mathcal{M}$\\
$N_{WC}$ & number of columns in  $\mathcal{W}$\\
$N_{SA}$ & simulated annealing steps \\
\hline
\end{tabular}
\caption{Main abbreviations and mathematical symbols adopted in this work.}
\end{table}

\subsection{Universal Set of Quantum Logic Gates}
\label{appendix:gates}

In this work, we employ a universal set of quantum logic gates, which are fundamental building blocks for quantum computation. 
These gates enable the manipulation of qubits and the implementation of complex quantum algorithms. 
The gates used in this study are summarized in Table~\ref{table:quantum_gates}.

\begin{table}[h]
    \centering
    \caption{Universal Set of Quantum Logic Gates}
    \label{table:quantum_gates}
    \begin{tabular}{|c|c|}
        \hline
        \textbf{Gate} & \textbf{Matrix Representation} \\ \hline
        $X$ & $\begin{pmatrix} 0 & 1 \\ 1 & 0 \end{pmatrix}$ \\ \hline
        $Y$ & $\begin{pmatrix} 0 & -i \\ i & 0 \end{pmatrix}$  \\ \hline
        $Z$ & $\begin{pmatrix} 1 & 0 \\ 0 & -1 \end{pmatrix}$ \\ \hline
        $H$ & $\tfrac{1}{\sqrt{2}} \begin{pmatrix} 1 & 1 \\ 1 & -1 \end{pmatrix}$  \\ \hline
        $R_z(\theta)$ & $\begin{pmatrix} e^{-i\theta/2} & 0 \\ 0 & e^{i\theta/2} \end{pmatrix}$  \\ \hline
        $T$ & $\begin{pmatrix} 1 & 0 \\ 0 & e^{i\pi/4} \end{pmatrix}$ \\ \hline
        $T/2$ & $\begin{pmatrix} 1 & 0 \\ 0 & e^{i\pi/8} \end{pmatrix}$ \\ \hline
        $C_X$ & $\begin{pmatrix} 1 & 0 & 0 & 0 \\ 0 & 1 & 0 & 0 \\ 0 & 0 & 0 & 1 \\ 0 & 0 & 1 & 0 \end{pmatrix}$ \\ \hline
    \end{tabular}
\end{table}

\section{Network science analysis}

\subsection{Mathematical structure of the CNOT graph}
The structure of the maze is generated by randomly connecting a control qubit $i$ with a target qubit $j$ for $1 \leq i,j \leq n$. This is equivalent to generating a random graph, where the connections (the CNOTs) are drawn uniformly over the set of possible connections between available vertices (the circuit wires). Since the sampling is performed with replacement, the graph that we generate is effectively a Erd\H{o}s--R\'enyi multigraph \cite{erdos59a, barabasi2016network} with $N_c$ edges (CNOT gates) and $N_q$ vertices (wires). The average degree $\langle k \rangle$ of the graph is given by:
\begin{align}
    \langle k \rangle = \frac{2N_c}{N_q},
\end{align}
and the degree distribution is Poissonian \cite{erdos59a}:
\begin{align}
    P_k = \langle k \rangle \frac{e^{- \langle k \rangle}}{k!}.
\end{align}
The Erd\H{o}s--R\'enyi graph exhibits a phase transition between $\langle k \rangle < 1$ and $\langle k \rangle > 1$. This phase transition corresponds to a change in the graph structure: for $\langle k \rangle < 1$, the graph is composed of scattered, isolated components, but in the regime in which $\langle k \rangle \geq 1$ a single, so-called giant component emerges. This behaviour corresponds to the phase transition between the traversable maze and the non-traversable maze, before the SA is considered. In fact, the use of the SA algorithm alters the structure of the graph, so that the resulting degree distribution does not correspond anymore to the one of the Erd\H{o}s--R\'enyi graph -- see Figure~\ref{fig:Centralities}. This is also reflected in the simulations that make use of opinion dynamics -- see Figure~\ref{fig:GMM}.

\subsection{Analytical derivation of walls distribution before simulated annealing effect} \label{appendix:p_before_annealing}

\begin{figure*}[t]
    \centering
    \includegraphics[width=0.75\textwidth]{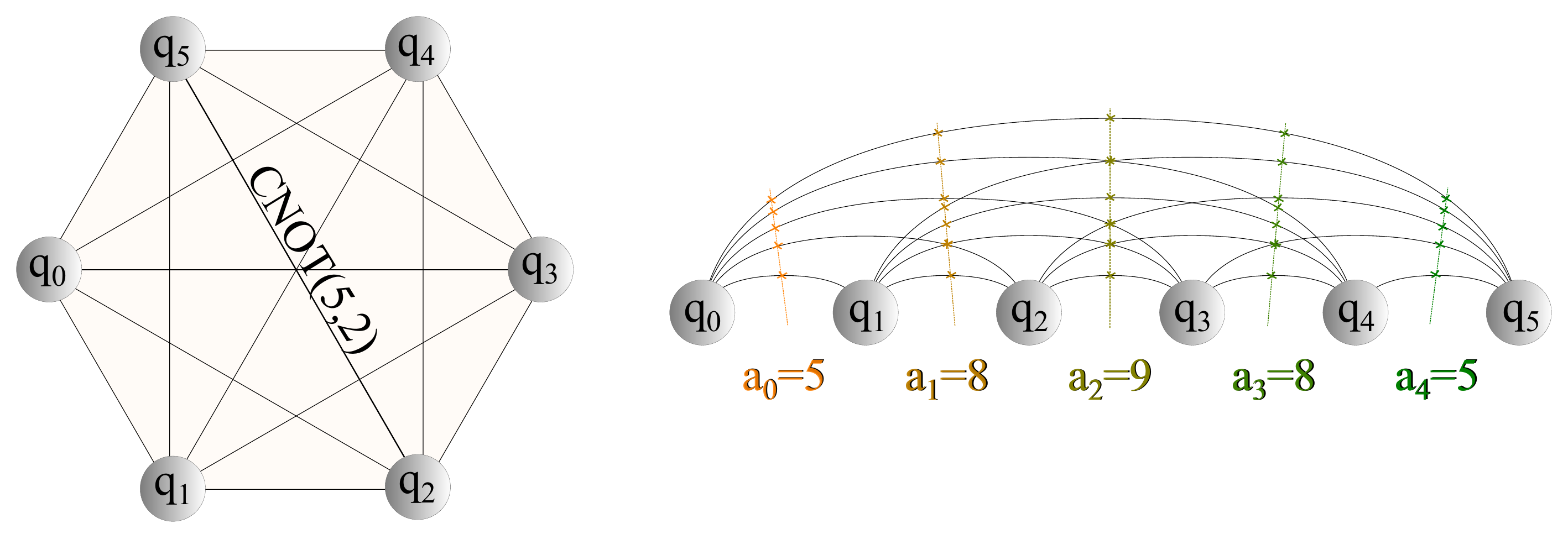}
    \caption{\raggedright The figure shows how randomly distributed CNOT gates can be thought of as links between six qubits arranged on a circle.
    Averaging over many circuits, the graph is effectively fully connected.
    After flattening the circle, the number of CNOT edges crossed by a vertical line between two nodes $q_i$ and $q_j$ indicates the average number of covered cells in each row of the $\mathcal {M}$ representation.
    For a maze with $N$ rows, the average number of covered cells in row $k$ is $a_k = -k^2 + (N-1)k + N$.}
    \label{fig:cnot count}
\end{figure*}

CNOT gates are sampled uniformly across qubits. 
Consequently, before SA is applied, the graph representing the CNOT circuit is, on average, fully connected, as depicted in Figure~\ref{fig:cnot count}. 
This observation can be exploited to derive an analytical expression for the distribution of the maze walls.

Let us assume that the circuit contains $N+1$ qubits. 
Therefore its corresponding maze has $N$ rows, indexed by $k=0, ..., N-1$. 
The average number of covered cells in row $k$ is given by:
$$\displaystyle a_k = -k^2 + (N-1)k + N. $$
To obtain a normalized distribution, we impose
$$\mathcal{A} = \sum_{k=0}^{N-1} a_{k} = 1\Longrightarrow \mathcal{A} = \frac{N(N+1)(N+2)}{6}.$$
Therefore, the normalized number of cells covered in row $k$ becomes
$$\displaystyle a_k = \frac{6[-k^2 + (N-1)k + N]}{N(N+1)(N+2)}.$$
Setting $n=N-1$ so that $k=0,...,n$, we can rewrite this expression as
$$\displaystyle a_k = \frac{6 [k(n-k) + n + 1]}{(n+1)(n+2)(n+3)}.$$
To study the limit distribution, we introduce the re-scaled continuous variable
$$\displaystyle y= \frac{k}{n}\in [0,1] \quad\text{so that}\quad k= nx.$$
and insert it into the previous expression to obtain:
$$\displaystyle a_n(y) = \frac{6[y n(n-ny) + n+1]}{(n+1)(n+2)(n+3)}$$
$$a_n(y)\sim \frac{6[n^2 (y-y^2) + \mathcal{O}(n)]}{n^3 + \mathcal{O}(n^2)}.$$
To compute the continuous density associated with the interval width $\delta y = \frac{1}{n}$, we multiply by $n$:
$$\displaystyle p_n(y) = na_n(y) = \frac{6[n^3 (y-y^2) + \mathcal{O}(n^2)]}{n^3 + \mathcal{O}(n^2)}.$$
Therefore, in the continuous limit the distribution converges to
$$\displaystyle p_n(y) \underset{n \to \infty}{\longrightarrow}  6y(1-y) \quad\text{where}\,\, \int_0^1 dy\, p(y) = 1,$$
which is a $\text{Beta}(y; 2,2)$ distribution.

\subsection{Interpretation of the cost function}
We assume that the maze is described by a binary matrix $\mathcal{M} \sim \mathcal{P}$. The binary matrix corresponds to the random sampling of CNOT gates between qubits $1\leq i,j \leq N_q$ and for a number of CNOTs $N_q$, described by the probability distribution $\mathcal{P}$. We assume that each row of the maze contains only one CNOT gate. We start from the cost function:
\begin{align}
    \mathcal{C} = \sum_{j=1}^{N_c} \sum_{i=1}^{N_q} \omega_i m_{ij},
\end{align}
where $m_{ij}$ are binary variables describing the position of the walls of the corresponding cells in a $N_q \times N_c$ grid and the (non-renormalized) Gaussian-like weights $\omega_i$ are given by:
\begin{align}
\omega_i = \exp{- \frac{N_q^2}{2\sigma^2 }\left(\frac{i}{N_q} - \frac{1}{2}\right)^2}.
\end{align}
As the weights $\omega_i$ only depend on the horizontal coordinate, i.e., the qubit wire, we can define the variable:
\begin{align}
    h_i = \sum_{j=1}^{N_c} m_{ij}.
\end{align}
and rewrite the cost function as:
\begin{align}\label{eq:cost_h_i}
    \mathcal{C} = \sum_{i=1}^{N_q} \omega_i h_i.
\end{align}
The random variable $h_i$ for a sampled configuration of the maze $\mathcal{M} \sim \mathcal{P}$ accounts for the number of walls in a single row of the maze. It also represents the $i$th bin in the histogram of the probability distribution $\mathcal{P}$.
For each column $k$ in the maze, there is only one sampled connection between qubits $i_k, j_k$ for $k=1,...,N_c$. As a result, we can also write the cost function $\mathcal{C}$ using the maze coordinates $Y_i,\ i=1,...,N_q$ -- horizontal coordinates can be omitted since CNOTs are only connected vertically -- as:
\begin{align}
    \mathcal{C} = \sum_{k=1}^{N_q} \dist^{\omega}(Y_{i_k}, Y_{j_k}),
\end{align}
where the weighted distance $\dist^\omega(\cdot, \cdot)$ is given by:
\begin{align}
    \dist^{\omega}(Y_{i}, Y_{j}) = \sum_{r=\min(Y_i, Y_j)}^{\max(Y_i,Y_j)-1} \omega_r.
\end{align}

\begin{figure*}
    \centering
    \includegraphics[width=0.9\linewidth]{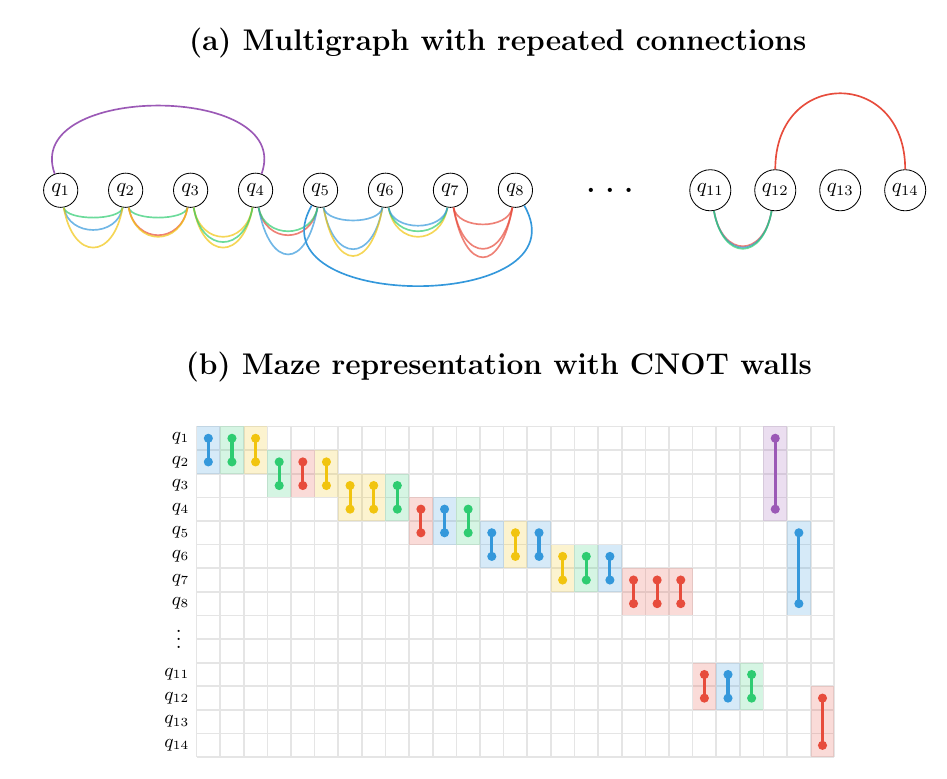}
    \caption{\raggedright(a) An example of a multigraph with $n=14$ nodes $q_1,...,q_{14}$ (number of wires or qubits) and repeated pairwise connections between the vertices (number of CNOTs or other entangling gates). (b) The maze representation of the multigraph in (a) with the corresponding connections represented in each column as walls of different colors matching the colors used for the edges in (a). The vertical axis corresponds to the number of wires, whereas the horizontal axis matches the total number of edges, including repetitions.}
    \label{fig:multigraph_and_maze}
\end{figure*}
The same cost function can also be written in a more compact way. The procedure of sampling one CNOT per row of the maze is equivalent to sampling multiple pairwise connections between vertices in a graph, with repetition. As the graph has multiple connections per pair, it is, more specifically, a multigraph. Assuming the adjacency matrix -- which can now have as entries integers larger than one to account for the multiple connections -- is given by $A$ and sampled randomly with repetition as described above, the cost function over the random graph structure can now be written as:
\begin{align}
    \mathcal{C} = \sum_{1 \leq i, j \leq N_q} A_{ij} \dist^{\omega}(Y_{i}, Y_{j}).
\end{align}
To summarize, our maze cost function can be mapped to a graph cost function typically implemented in community detection \cite{Bullo2018}. 
A particular case is the one in which $\omega_i = 1$. In this case, the function $\dist^{\omega}(Y_{i}, Y_{j})$ reduces to the $L_1$ distance, i.e., $\dist^{\omega=1}(Y_{i}, Y_{j}) = | Y_{i} - Y_{j} |$. Therefore, the cost function reduces to:
\begin{align}
    \mathcal{C}[\omega=1] = \sum_{1 \leq i, j \leq N_q} A_{ij} | Y_{i} - Y_{j} |.
\end{align}

\subsection{A simplified coarse-graining procedure of the maze}
We want now to analyze the solution provided by the SA algorithm in the limit of large numbers of qubits $N_q\gg1$. This is the same limit used to derive the asymptotic profile of the wall distribution, i.e., the Beta distribution $\text{Beta}(y,2,2) = h_0(y) = 6y(1-y)$. We use then a coarse-graining procedure on the cost function, by rewriting the sums in terms of integrals over average densities \cite{Espanol2004} in the continuous limit:
\begin{align}
    \frac{1}{N_q} \sum_{i} a_i (...) \approx \int \rho(y) a(y) ... dy 
\end{align}

For $N_q \gg 1$, we approximate the energy as:

\begin{align}\label{eq:C_omega_x}
    \mathcal{C} \approx   \int_0^1 V_{\text{eff}}(y) p(y) dx,
\end{align}

where $V_{\text{eff}}(y)$ is an effective potential.

\subsection{Boundary conditions}
In Section \ref{appendix:p_before_annealing} we derived the analytical shape of the wall distribution, which in the limit of infinite wires should correspond to a $\text{Beta}(y,2,2)$ distribution. However, for the histogram of the wall distribution in Figure~\ref{fig:simulated annealing}, we notice that CNOT wall counts are non-zero at the boundaries $y=0$ and $y=1$. To account for this finite-scaling effect, we introduce a slightly modified shape of the analytical distribution which allows us to better describe the effective behaviour of the distribution for $N_q<\infty$. More specifically, we start from the theoretical continuous limit of the CNOT wall histogram:
\begin{align}
    p_0(y) = 6y(1-y),
\end{align}
and introduce the following modified functional form:
\begin{align}
    \tilde{p}_0(y) = b(y) + \frac{M}{Z} p_0(y),
\end{align}
where $b: \mathbb{R} \mapsto \mathbb{R}_{>0}$ is a boundary function and where 
\begin{align}
    Z = \int_0^1 p_0(y) dy \\ 
    M = 1 - \int_0^1 b(y) dy \label{eq:M_b}
\end{align}
are normalization constants. From the histogram shown in Figure~\ref{fig:simulated annealing}, we can observe two properties of the modified distribution: (a) the distribution is not zero at the boundaries, but evaluates to a value $c\geq0$; (b) the distribution has a vertical offset $d \geq 0$ with respect to $p_0$, (c) the  function $\tilde{p}_0$ is continuous, so we need to ensure the overall continuity at the interface between $p_0$ and $b$. We therefore suggest the following functional shape for $b$:
\begin{align}
    b(y) = d + (c-d)\left(\frac{e^{-y/\delta} + e^{-(1-y)/\delta}}{1 + e^{-\delta}} \right),
\end{align}
which resembles a box function for $\delta \sim 0$, but due to its smoothness preserves the continuity of the distribution $\tilde{p}_0$.

\subsection{Coarse-Grained Free Energy}
We use here a procedure commonly employed in coarse-graining approaches, see, e.g.,\cite{Espanol2004, CastiglioneCoarseGraining}. The goal of the SA algorithm is to find the probability distribution that minimizes the cost function $\mathcal{C}$. During the minimization, the total number of CNOTs (edges) is invariant, so that we can treat this problem as a probability transport task. We use Eq.~\eqref{eq:C_omega_x} for the cost function and add the relative entropy, or KL divergence  -- see also Ref.~\cite{Thaler_2022} --, which is often employed to model the transition from a many-body potential to a coarse-grained mean-field potential. We assume here that the SA algorithm does not modify the overall shape of the boundary function $b(y)$ -- this seems to be confirmed by our simulations, although the SA procedure seems to alter the value of the two offsets $c, d$ contained in $b(y)$. As a result, the SA algorithm only alters the functional form of $p_0$. \\
The coarse-grained free energy $F_T[p]$ for $n\gg1$ is given by:
\begin{align}
F_T[p] = \int_0^1 V_{\text{eff}}(y) p(y) \, dy + T S[p_0, p],
\end{align}
where
\begin{align}
    S[p_0, p] = \int_0^1 p(y) \log \left( \frac{p(y)}{p_0(y)} \right) \ dy
\end{align}
is the relative entropy between $p_0$ and $p$.
The minimization procedure is subject to the constraint:
\begin{align}\label{eq:prob_constraint_min}
\int_0^1 p(y) \, dy = 1,
\end{align}
due to the conservation of probability. If the histogram not renormalized, the value of the integral can be set equal to a constant. The cost functional including a Lagrange multiplier for the constraint in Eq.~\eqref{eq:prob_constraint_min} is given by:
\begin{align}\label{eq:Lambda_var_min}
    \mathcal{L}[p, \lambda] = F_T[p] + \lambda \left(\int_0^1 p(y) dy - 1\right)
\end{align}

The variational minimization of $\mathcal{L}$ in Eq.~\eqref{eq:Lambda_var_min}:
\begin{align}
    \frac{\delta \mathcal{L}}{\delta p} = 0,
\end{align}
leads to
\begin{align}
 V_{\text{eff}}(y) + T \left(\log \frac{p(y)}{p_0(y)} + 1\right) = -\lambda,
\end{align}
where $\lambda$ is a Lagrange multiplier. Solving for $p(y)$ gives:
\begin{align}
p_T^*(y) = \frac{1}{Z} p_0(y) \exp\left(-\frac{V_{\text{eff}}(y)}{T}\right), \\
\quad Z^* = \int_0^1 p_0(y) e^{- V_{\text{eff}}(y)/T} \, dy.
\end{align}
The new distribution that includes boundary value effects is then given by:

\begin{align}\label{eq:boundary_exp_fit_model}
    \tilde{p}_T^*(y) = b(y) + \frac{M}{Z^*} p_T^*(y) 
\end{align}

\subsection{Shape of the effective potential}\label{sec:effective_potential}

The negative effective potential should be peaked at the center of the qubit register, i.e., around $y=1/2$, to model the repulsive interaction between the two clusters of CNOTs. It should also be minimal at two values to the boundaries where the clusters will be located. 

For such a functional profile in $y$, as $T$ approaches zero, the damping exponential term strongly suppresses the center ($y = 1/2$), since the potential is maximal there. This results in a bimodal distribution with peaks near the boundaries ($y_1 = \mu$ and $y_2 = 1 - \mu$), consistent with the observed behavior of two clusters migrating to the sides during simulated annealing.

Let us consider the function $\mathcal{C}$ as defined in Eq.~\eqref{eq:cost_h_i}:

\begin{align}\label{eq:C_Veff_x}
    \mathcal{C}(y) = \sum_i h_i \omega_i \approx N_q \int_{0}^1 \omega(y) p(y) dy,
\end{align}
where $\omega(y)$ is the continuous version of the $\omega_i$ weights and $y = i/N_q$:
\begin{align}
    \omega(y) = \frac{1}{\sqrt{2\sigma^2\pi}}\exp{-\frac{N_q^2}{2\sigma^2}\left(y - \frac{1}{2}\right)^2}.
\end{align}

As previously highlighted, the effective potential should have a peak in the middle of the $y$ axis, which is the interaction that the values $\omega_i$ effectively generate. To better model a diffusion-like behaviour during the Simulated Annealing procedure, we assume that $\sigma \sim \sqrt{N_q}$
However, in principle, when we choose the fit profile, any parametric potential with a peak at $y=\frac{1}{2}$ can be a suitable candidate. A good family of candidates is given by polynomials with a maximum at $y=1/2$ whose values disappear at the boundaries, i.e., $p(0) = p(1) = 0$:
\begin{align}
    p_n(y) = A_n y^n(1-y)^n.
\end{align}
and
\begin{align}
    p_n(y + 1/2) = A_n (y - 1/2)^n
\end{align}
These are symmetric Bernstein polynomials. For instance, an infinite sum of such polynomials:
\begin{align}
    \sum_{n=0}^N c_n p_n(y)
\end{align}
generates the exponential $\omega(y)$ for $c_n = 1/n!$:
\begin{align}
    \omega(y) = \sum_{n=0}^{\infty} \frac{(-1)^n}{n!} p_{2n}(y+1/2).
\end{align}

The fit model has limitations due to its underlying assumptions. First, it assumes that the coarse-grained cost function can be written down as the mean value of the Gaussian weight $\omega(y)$ over a probability distribution function $p(y)$. In order to study the dynamical properties of these graph, we turn to out-of-equilibrium algorithms widely used in network science that allow us to study the propagation of information within the graph and the emergence of clusters. 

\begin{figure}
    \centering
    \hspace{-0.5cm}
    \includegraphics[width=0.95\linewidth]{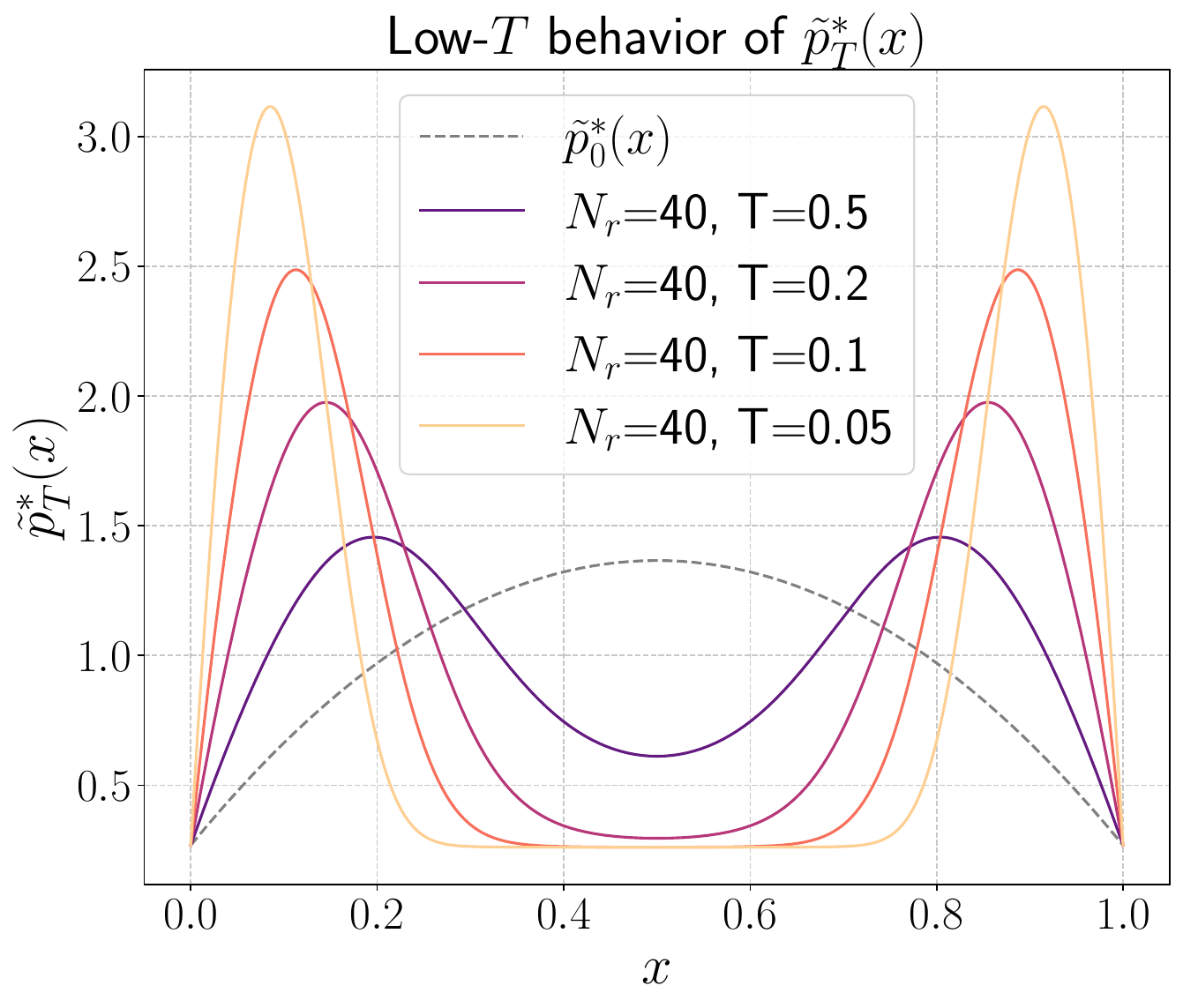}
    \caption{\raggedright Visualization of the evolution of the ansatz $\tilde{p}^*(x)$ -- including the boundary function $b(x)$ --with decreasing temperature $T$. We can see that the shape correctly describes the separation of the distribution into two clusters. The initial profile of $\tilde{p}_0(x)$ is shown as a dotted gray line. The values of the parameters have been set to $d=0.5$, $c=0.02$, $l=0.1$, $\sigma=\sqrt{N_q}$}
    \label{fig:enter-label}
\end{figure}

Despite its limitations, this model seems to correctly reproduce the dynamics of the SA separation. We can see in Figure~\ref{fig:distribution fit} that the functional profile given in Eq.~\eqref{eq:boundary_exp_fit_model} matches or surpasses the performance of the other models, in particular when the number of random CNOTs approaches the number of qubits in the circuit. For a low number of CNOTs, we expect the Erd\H{o}s--R\'enyi graph to consist on average of disconnected clusters, so a functional profile consisting of two separate Gaussian distribution shows a better correspondence with the data.

More specifically, we assume that when the two clusters generated by the ansatz given in Eq.~\eqref{eq:boundary_exp_fit_model} are sufficiently separated, they can be treated as two separate Gaussian distributions. Let us consider the separation potential
$V_{\text{eff}}(x)$: by definition, it should be peaked at $y=\frac{1}{2}$ to allow for the separation in two clusters and has two minima for $y_1=\mu$ and $y_2=1-\mu$, where the peaks of the two clusters are located. This configuration allows us to employ the Laplace method \cite{Butler2007} to approximate the partition function of the Gibbs distribution obtained in Eq.~\eqref{eq:boundary_exp_fit_model}, as $T \sim 0$. In the case of two equal maxima, we need to apply the Laplace method on each maximum and sum the contributions \cite{majdalani_laplace_method}. The Laplace approximation for $Z^*$ gives:
\begin{align}
    &Z_i^* = \sqrt{\frac{2\pi T}{|V''_{\text{eff}}(y_i)|}} p_0(y_i) e^{-\frac{1}{T} V_{\text{eff}}(y_i)} \\
    &Z^* \approx Z_1 + Z_2
\end{align}

whereas for $p^*$ close to the maxima $y_1, y_2$ we can approximate the function using a second-order Taylor approximation on the exponential term (the envelope does not contribute significantly to the expansion):
\begin{align}
    p_i^*(y) \approx p_0(y_i) e^{-\frac{1}{T} V_{\text{eff}}(y_i) -  \frac{1}{2T} V''_{\text{eff}}(y_i) (y - y_i)^2},
\end{align}
where $i=1,2$.
If the two peaks are sufficiently distant, the overall shape is therefore given by the sum of these two contributions, since the others are suppressed in the limit of $T \sim 0$. The resulting distribution is thus a Gaussian mixture model (GMM) \cite{bishop2006pattern}:
\begin{align}
    p(y) \approx w_1 \frac{1}{Z_1^*}p^*_1(y) + w_2 \frac{1}{Z_2^*}p^*_2(y),
\end{align}
with $w_1 = Z^*_1/Z^*$ and $w_2 = Z^*_2/Z^*$. If the peaks are symmetric, then we must have $w_1 = w_2 = 1/2$. In our tests, this model performs well when the clusters are well separated and there are no overlap areas between them, which is also confirmed by our simulations of consensus dynamics -- see Figure~\ref{fig:GMM} --, where we also employ the same model. \\
We also test a third model that consists of two Beta distributions with an offset. This model assumes that the SA algorithm simply splits the original distribution into two separate but symmetric Beta distributions with modified coefficients $a, b$. Also this model proves effective, though it does not provide the same qualtitative description of the SA procedure as the model provided in Eq.~\eqref{eq:boundary_exp_fit_model}. Interestingly, it seems to be effective in the same regime as the model given in Eq.~\eqref{eq:boundary_exp_fit_model}.

\subsection{Construction of the graph for coarsed-grained approximation}
\label{app:graph_cons}
Starting from the Erd\H{o}s--R\'enyi model defined above, we extend the static graph representation to a time-resolved directed graph that encodes the circuit structure across its layers. In this construction, each node is associated with a qubit at a given circuit depth, and edges capture both intra-layer and inter-layer interactions induced by the gate sequence. This dynamical graph representation enables the implementation of consensus algorithms that mimic the propagation of correlations through the entanglement structure of the circuit, thereby providing a natural framework to study information flow and mixing across the two partitions.

To obtain a time-resolved representation of the circuit, we introduce the binary grid $A\in\{0,1\}^{n\times m}$, describing the multigraph in Figure~\ref{fig:multigraph_and_maze} defined by
\begin{equation}
A_{i\ell} = \mathbf{1}_{\{i=c_\ell\}} + \mathbf{1}_{\{i=t_\ell\}},
\qquad \sum_{i=1}^n A_{i\ell}=2 \ \ \forall \ell.
\end{equation}
We then define a directed graph $G_{rc}=(V_{rc},E_{rc})$ whose vertex set is
\begin{equation}
V_{rc} = \{(i,\ell)\in V\times\{1,\dots,m\} : A_{i\ell}=1\},
\end{equation}
with $|V_{rc}|=2m$. The edge set decomposes as $E_{rc}=E_{\mathrm{row}}\cup E_{\mathrm{col}}$, where
\begin{align}
E_{\mathrm{row}} &= \{ \big((i,\ell_r),(i,\ell_{r+1})\big) :
\ell_r,\ell_{r+1} \nonumber\\
&\quad\in T_i \text{ consecutive} \}, \\
E_{\mathrm{col}} &= \{ \big((c_\ell,\ell),(t_\ell,\ell)\big),
\big((t_\ell,\ell),(c_\ell,\ell)\big)\nonumber\\ &\quad:  \ell=1,\dots,m \},
\end{align}
and $T_i=\{\ell : A_{i\ell}=1\}$ denotes the set of times at which qubit $i$ participates in a gate. 

The graph $G_{rc}$ can therefore be interpreted as a layered, time-expanded representation of the underlying random multigraph $M$, in which each interaction event $e_\ell$ is lifted to a pair of time-stamped vertices and each qubit $i$ is associated with a directed chain linking its successive interaction events.

\subsection{Consensus Dynamics Model}
\label{app:Consensus}

Consensus dynamics describes how a collection of interacting agents can reach agreement through local interactions on a network.
Originally developed in control theory and distributed computation, consensus models are now widely used to study collective behavior in complex systems such as robotic swarms, sensor networks, social opinion dynamics, and synchronization phenomena.

Consider a network of $N$ agents represented by a graph $G=(V,E)$, where nodes correspond to agents and edges represent communication links.
Each agent $i$ holds a scalar state $x_i(t)$ that evolves through interactions with neighboring agents.
Let $\mathcal{N}_i$ denote the set of neighbors of node $i$.

A commonly studied continuous-time consensus protocol is

\begin{equation}
\dot{x}_i(t) =
\sum_{j \in \mathcal{N}_i}
a_{ij}\left[ x_j(t) - x_i(t)\right],
\end{equation}

where $a_{ij}$ denotes the interaction weight between agents $i$ and $j$.
In vector form the dynamics can be written as

\begin{equation}
\dot{\mathbf{x}}(t) = -L \mathbf{x}(t),
\end{equation}

where $L = D - A$ is the graph Laplacian, with $A$ the adjacency matrix and $D$ the degree matrix.
The Laplacian structure implies that consensus dynamics can be interpreted as a diffusion process on the graph.

When the graph is connected the system converges to a common value shared by all agents.
The convergence rate is governed by the spectral properties of the Laplacian matrix.
In particular, the second-smallest eigenvalue $\lambda_2(L)$, known as the \emph{algebraic connectivity} or \emph{Fiedler value}, determines the speed of consensus formation.

Consensus processes also admit a discrete-time formulation known as the \emph{DeGroot model}, in which agents repeatedly average the states of their neighbors:

\begin{equation}
\mathbf{x}(t+1) = W \mathbf{x}(t),
\end{equation}

where $W$ is a row-stochastic interaction matrix encoding the influence of neighboring nodes.

In the context of the present work, the nodes of the interaction graph correspond to the CNOT events of the circuit, two per CNOT gate, one on the control wire and one on the target — while edges are of two types: bidirectional edges joining the two nodes of each CNOT, encoding the entangling interaction, and directed edges joining successive events along the same qubit wire, encoding the temporal order of operations. The consensus variable $x_i(t)$ can therefore be interpreted as a scalar quantity attached to each entangling event, which propagates through the circuit as interactions couple different qubit wires.
Under this interpretation the consensus dynamics provides a simple model of how correlations spread along the time-ordered interaction structure generated by the sequence of CNOT gates.

Under this interpretation the consensus dynamics provides a simple model of information spreading along the directed interaction graph generated by the sequence of CNOT gates.
If the circuit naturally separates into weakly interacting regions, the dynamics tends to form clusters of nodes whose values remain separated.
Conversely, if many interactions connect the regions, the consensus process mixes the values across the network.

For numerical stability and simplicity we employ a discrete-time update rule in which each node retains a fraction of its current value while incorporating the average value of its outgoing neighbors.
The update rule used in our simulations is therefore

\begin{equation}
x_i(t+1) =
(1-\epsilon)x_i(t)
+
\epsilon
\frac{1}{|\mathcal{N}_i^{\mathrm{out}}|}
\sum_{j\in \mathcal{N}_i^{\mathrm{out}}} x_j(t),
\end{equation}

where $\epsilon \in [0,1]$ (chosen as $0.2$) controls the strength of the interaction and $\mathcal{N}_i^{\mathrm{out}}$ denotes the set of nodes connected to node $i$ through outgoing edges.
This corresponds to a normalized discrete-time consensus process on the directed interaction graph induced by the circuit.

The initial condition for the consensus dynamics is then chosen as
\begin{equation}
x_i(0) =
\begin{cases}
\xi_i^{+} \in [0,1], & i \in V_A, \\
\xi_i^{-} \in [-1,0], & i \in V_B,
\end{cases}
\end{equation}
\label{eq:init_cond_cons}
with $\xi_i^{\pm}\) random variables. The initial condition in Eq.\ref{eq:init_cond_cons} is chosen to encode the ideal bipartition: nodes in the upper half are seeded with positive values and those in the lower half with negative ones, representing a perfectly separated configuration with respect to the cut identified by the maze agent. The consensus dynamics then measures how far the circuit departs from this ideal. If the cut induces a genuine separation, the seeded configuration is a near-stationary state of the dynamics and the bimodal structure is preserved; if the two halves are in fact coupled by CNOT gates crossing the cut, the dynamics transports values across the interface and the seeded separation degrades. The overlap area therefore quantifies the deviation of the actual circuit from an idealized, perfectly partitioned one — it measures how much the partition fails, not whether a partition spontaneously emerges.

\begin{figure}
    \centering
    \includegraphics[width=\linewidth]{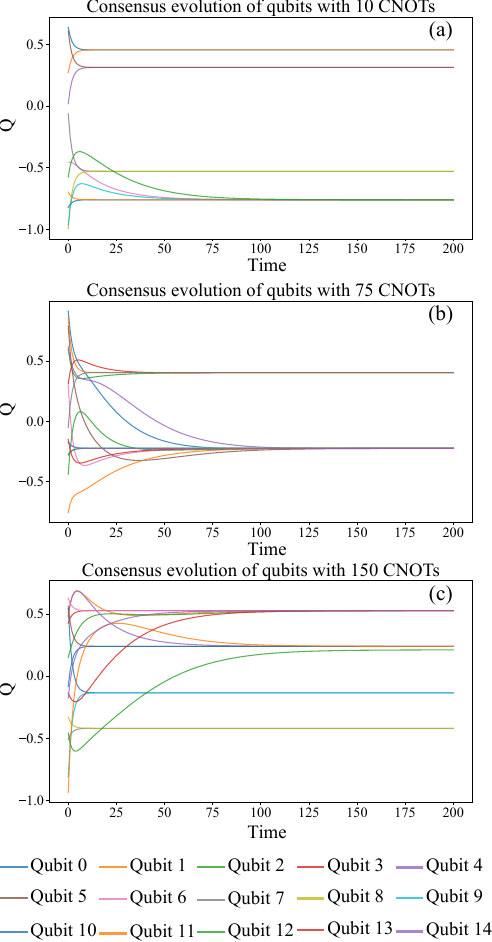}
    \caption{\raggedright Single-trajectory consensus dynamics of qubit values for circuits with different numbers of CNOT gates. 
    (a) For a small number of CNOTs (10), the qubits form two well-separated clusters that remain distinct throughout the evolution. 
    (b) For an intermediate number of CNOTs, the clusters begin to mix as interactions cross the ideal partition. 
    (c) For a large number of CNOTs, strong mixing occurs and the final qubit values converge, preventing a clear clustering. 
    This indicates that the circuit can no longer be cleanly partitioned into two subcircuits.}
    \label{fig:consensus_traj_examples}
\end{figure}

Figure~\ref{fig:consensus_traj_examples} shows three representative trajectories of the consensus dynamics for circuits with different numbers of CNOT gates, in particular we consider $Q$ as the value of the last node for each qubit line. 
Each trajectory illustrates how the qubit values evolve over time under the update rule introduced in the previous section. 
These examples highlight how the structure of the interaction graph induced by the circuit determines the propagation of information across the qubits.\\

\textbf{Low interaction regime.} In Figure~\ref{fig:consensus_traj_examples}(a), corresponding to a circuit containing only 10 CNOT gates, the qubits naturally separate into two distinct groups from the very beginning of the evolution.
The consensus dynamics propagates information only within each group, while interactions between the two groups are essentially absent.
As a consequence, the qubit values converge to two clearly separated clusters, approximately located around positive and negative values.
The final state of the system therefore remains easily partitionable into two sets of qubits with opposite signs.\\
\textbf{Intermidiate interaction regime.} In Figure~\ref{fig:consensus_traj_examples}(b), corresponding to a circuit with 75 CNOT gates, the situation changes significantly.
While the system still starts from two distinct clusters, the qubit values begin to mix shortly after the beginning of the dynamics.
This behavior indicates the presence of CNOT interactions connecting the two previously separated regions of the circuit.
As a result, the final qubit values become closer across the two groups, and the separation between clusters becomes less pronounced. \\
\textbf{High interaction regime.} Finally, Figure~\ref{fig:consensus_traj_examples}(c) shows the evolution for a circuit containing 150 CNOT gates.
In this regime the interaction graph becomes sufficiently dense that information propagates across the entire network of qubits.
The consensus dynamics rapidly mixes the values of all nodes, and the final configuration no longer exhibits any clear clustering structure.
This behavior indicates that the circuit cannot be partitioned into two weakly interacting regions, and therefore the simulated annealing procedure cannot identify a meaningful split of the circuit.

These examples illustrate how consensus dynamics provides a simple dynamical probe of the interaction structure of the circuit: well-separated regions of the circuit manifest themselves as persistent clusters in the consensus trajectories, whereas strong interconnections lead to rapid mixing and loss of cluster structure.

\subsection{Gaussian Mixture Model}
\label{app:GMM}
To quantify whether the consensus trajectories exhibit a clustered structure, we employ a Gaussian Mixture Model (GMM) to identify groups of qubits with similar final values. 
This allows us to automatically determine whether the qubits separate into two distinct sets after the consensus dynamics has converged.

Let $x_i(T)$ denote the final value of qubit $i$ after the consensus evolution described in the previous section.
The dataset therefore consists of the scalar values

\begin{equation}
\mathcal{X} = \{x_1(T), x_2(T), \dots, x_N(T)\}.
\end{equation}

We model this distribution as a mixture of two Gaussian components

\begin{equation}
p(x) =
\sum_{k=1}^{2}
\pi_k \, \mathcal{N}(x \mid \mu_k, \sigma_k^2),
\end{equation}

where $\pi_k$ are the mixture weights satisfying $\sum_k \pi_k = 1$, and $\mathcal{N}(x \mid \mu_k,\sigma_k^2)$ denotes a Gaussian distribution with mean $\mu_k$ and variance $\sigma_k^2$.

The parameters $(\pi_k, \mu_k, \sigma_k)$ are estimated using the Expectation Maximization (EM) algorithm (Alg. \ref{alg:GMM}).
Once the model is fitted, each qubit value $x_i(T)$ is assigned to the component with the highest posterior probability.

The resulting cluster assignments provide a quantitative indicator of whether the circuit can be separated into two weakly interacting regions.
When two well-separated Gaussian components are identified, the qubits naturally divide into two groups.
Conversely, when the two components strongly overlap, the circuit exhibits strong mixing and no clear partition emerges.

\begin{algorithm}[H]
\caption{Gaussian Mixture Clustering of Consensus Values}
\begin{algorithmic}[1]

\Require Final consensus values $\{x_1(T),\dots,x_N(T)\}$

\State Initialize mixture parameters $(\pi_k,\mu_k,\sigma_k)$ for $k=1,2$

\Repeat

\State \textbf{E-step:} compute posterior responsibilities
\begin{equation*}
\gamma_{ik} =
\frac{\pi_k \mathcal{N}(x_i|\mu_k,\sigma_k^2)}
{\sum_{j=1}^{2} \pi_j \mathcal{N}(x_i|\mu_j,\sigma_j^2)}
\end{equation*}

\State \textbf{M-step:} update parameters
\begin{equation*}
\mu_k =
\frac{\sum_i \gamma_{ik} x_i}{\sum_i \gamma_{ik}}
\end{equation*}

\begin{equation*}
\sigma_k^2 =
\frac{\sum_i \gamma_{ik}(x_i-\mu_k)^2}{\sum_i \gamma_{ik}}
\end{equation*}

\begin{equation*}
\pi_k =
\frac{1}{N}\sum_i \gamma_{ik}
\end{equation*}

\Until{convergence}

\State Assign each qubit to cluster $k = \arg\max_k \gamma_{ik}$

\end{algorithmic}
\end{algorithm}
\label{alg:GMM}

The clustering results obtained through this procedure allow us to determine whether the consensus dynamics preserves two separated groups of qubits or whether the values become fully mixed. In particular, circuits that can be partitioned into two weakly interacting regions tend to produce two well-separated Gaussian components, whereas strongly connected circuits lead to overlapping distributions and therefore poor cluster separability.

\begin{figure}
    \centering
    \includegraphics[width=\linewidth]{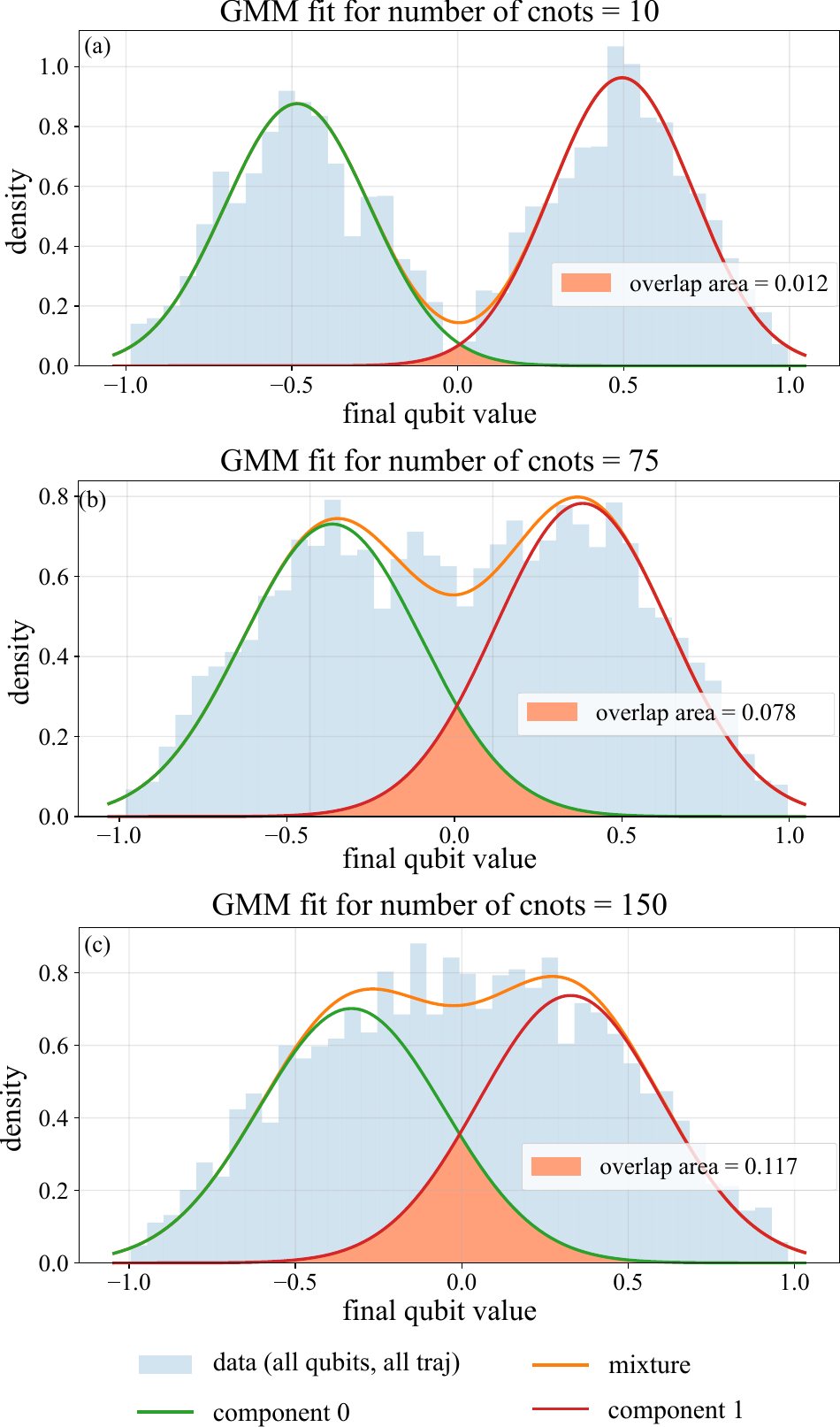}
    \caption{\raggedright Gaussian Mixture Model (GMM) analysis of the final consensus values $x_i(T)$ for 30 qubits over 500 trajectories, with increasing circuit connectivity: (a) 10, (b) 75, and (c) 150 CNOTs gates.
    Blue histograms represent the empirical distribution of all qubit values, while the solid curves correspond to the fitted two-component Gaussian mixture and its individual components.
    The shaded region denotes the overlap between the two Gaussian distributions and provides a quantitative measure of cluster separability.
    In the low-interaction regime (a), the overlap is minimal, indicating that qubits remain clearly partitioned into two distinct groups.
    As the number of CNOTs increases (b,c), the overlap grows substantially, reflecting enhanced mixing between qubits and the progressive breakdown of a two-cluster structure.
    This behavior highlights the limitations of the simulated annealing procedure in identifying clean bipartitions in highly connected circuits.
    }
    \label{fig:GMM}
\end{figure}

In the following plots (Figure~\ref{fig:GMM}) we run 500 different trajectories with 2500 SA optimization steps for 30 qubits with respectively 10, 75 and 150 CNOTs.

We notice that in Figure~\ref{fig:GMM}(a) the average overlap over 500 trajectory is very very small sign that we are indeed in the Low interaction regime. Increasing the number of CNOTs we can clearly see that the overlap area between the gaussians fitting the dataset increases (Figure~\ref{fig:GMM}(b)(c)). This is a clear sign that the SA algorithm struggles to split the circuit in two ideal partitions.

\subsection{Conductance}
\label{app:Conductance}
Another measure that we can be used in order to define how connected are our two ideal partitions is the conductance. If we consider the undirected version of the graph that represents our quantum circuit as defined in the section \ref{sec:centralities}, we can define the two ideal partitions as the set of nodes beloging to the first half of the qubits in the quantum circuit (set $\mathcal{S}$) and the nodes related to the second half of the qubits (set $\hat{\mathcal{S}}$). In this setup we can define the conductance as:
\begin{equation}
    \varphi(S) = \frac{a(S, \hat{S})}{\text{min}(\text{vol}(S),\text{vol}(\hat{S}))}\quad,
\end{equation}
where
\begin{align*}
&a(S,T) = \sum_{i \in S}\sum_{j \in T} a_{ij} \text{,} \\
&\text{vol}(S) = a(S,V) = \sum_{i \in S}\sum_{j \in V} a_{ij}.    \quad 
\end{align*}

\begin{figure}
    \centering
    \includegraphics[width=\linewidth]{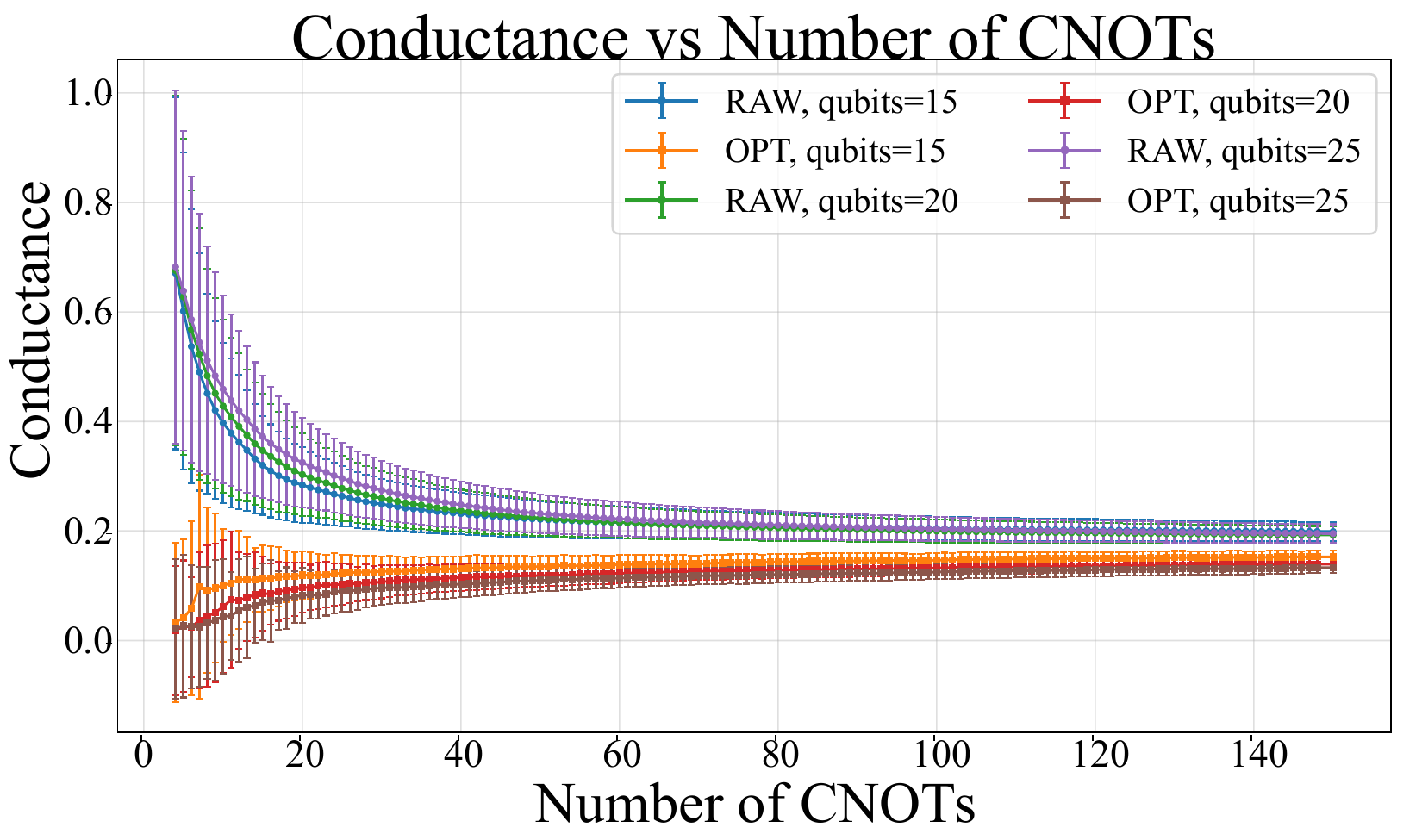}
    \caption{\raggedright Conductance as a function of the number of CNOT gates for circuits with different numbers of qubits.
        Solid lines represent the mean conductance over multiple realizations, while shaded regions (or error bars) indicate the standard deviation.
        For each system size, we compare the raw circuits (before optimization) with the optimized circuits obtained via simulated annealing.
        The optimization consistently reduces the conductance, indicating a suppression of edges connecting the two predefined partitions.
        At large CNOT densities, the conductance saturates to similar values for both cases, signaling the onset of a strongly mixed regime where no clear partition of the circuit can be identified.
        }
    \label{fig:conductance_vs_cnots}
\end{figure}

The results of this analysis are reported in Figure~\ref{fig:conductance_vs_cnots}.
We observe that the conductance of the raw circuits is initially large for small numbers of CNOTs, reflecting the presence of several connections between the two halves of the system.
As the number of CNOTs increases, the conductance decreases and approaches a plateau, indicating that the relative number of inter-partition edges becomes smaller compared to the minimum volume of the two partitions.

In contrast, the optimized circuits obtained via SA exhibit significantly lower conductance across the entire range of CNOTs.
This indicates that the optimization procedure effectively reduces the number of edges crossing the two partitions, promoting a separation of the circuit into two weakly interacting regions.

Interestingly, at large CNOT densities the conductance of both raw and optimized circuits converges to similar values.
This suggests that, beyond a certain connectivity threshold, the interaction graph becomes sufficiently dense that it is no longer possible to isolate two well-separated partitions.
This behavior is consistent with the loss of cluster structure observed in the consensus dynamics, and provides a complementary graph-theoretic signature of the mixing regime in the circuit.

\begin{figure*}[t]
    \centering
    \includegraphics[width=\textwidth]{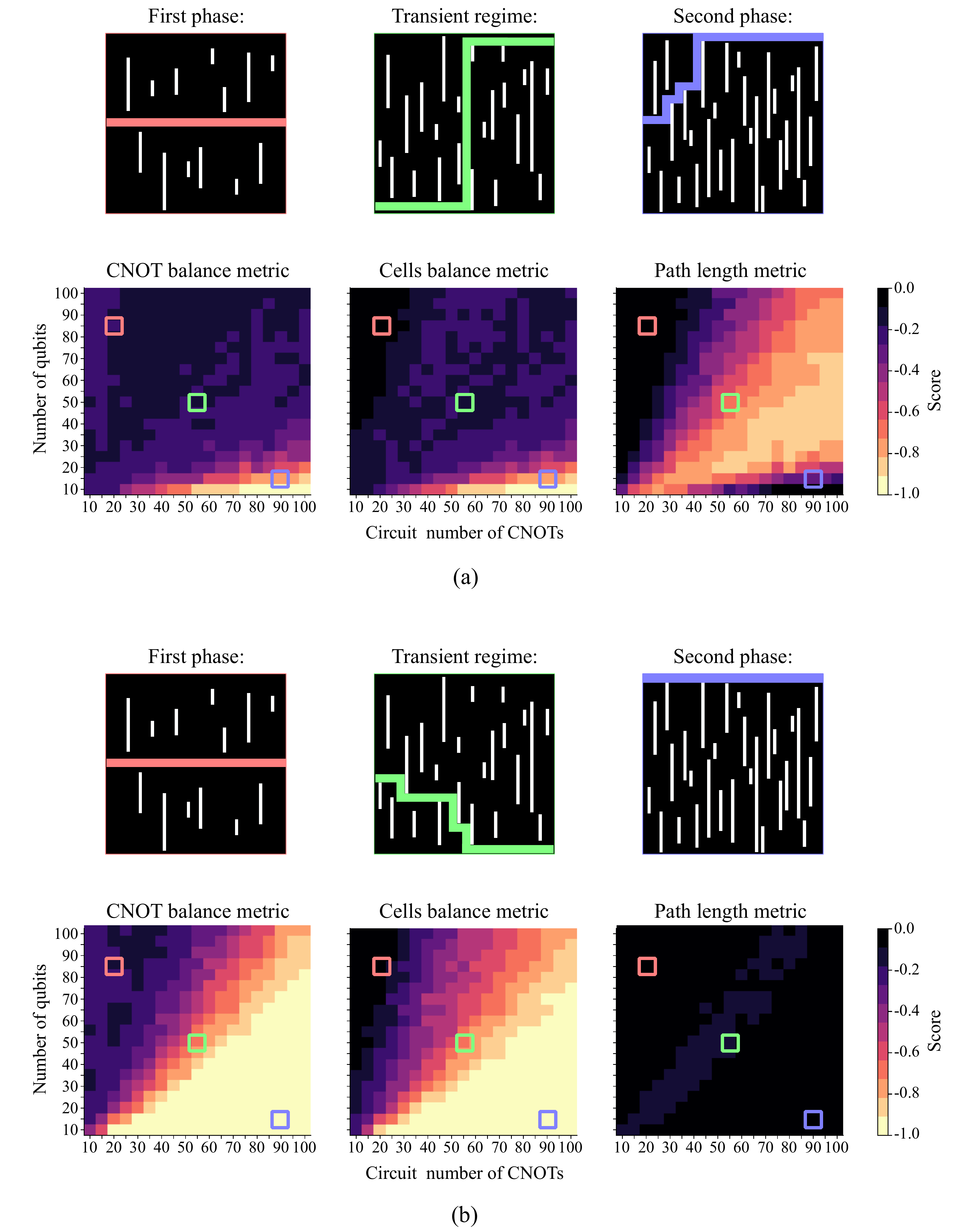}
    \caption{\raggedright The figure shows the score obtained by the best performing agent among all candidates at $\alpha=1, \beta=1, \gamma=1$ for the upper image (a) and $\alpha=1, \beta=1, \gamma=3$ for the lower picture (b). }
    \label{fig:gamma=1,3}
\end{figure*}

\subsection{Phase transition as a mass transport problem over graph}
\label{app:mass transport}

The phase transition shown in Figure~\ref{fig:maze metrics} can be analyzed within the consensus-dynamics framework introduced in Section~\ref{app:Consensus}. Rather than prescribing an \emph{a priori} bipartition of the circuit, we construct the initial condition directly from the partition identified by the agent in the maze representation.

Let $\mathcal{M}$ denote the maze associated with the circuit and let $\mathcal{C} \subset \mathcal{M}$ be the cut (best path) returned by the agent. This cut is first projected onto the wire (grid) representation via a map
\begin{equation}
\pi: \mathcal{M} \rightarrow \mathcal{W}_{\mathrm{grid}},
\end{equation}
which assigns to each circuit column a cut position across the qubit wires. The grid is then lifted to a directed graph $G=(V,E)$, built following the appendix section \ref{app:graph_cons}.

This induces a mapping
\begin{equation}
\phi: \mathcal{M} \rightarrow V,
\end{equation}
through which the cut defines a bipartition $V = V_A \cup V_B$, with $V_A \cap V_B = \varnothing$. Each node $v \in V$, associated with coordinates $(r,c)$, is assigned to $V_A$ or $V_B$ depending on its position relative to the projected cut. Boundary nodes are resolved through their bidirectional column connectivity, ensuring a consistent assignment across interacting qubits.

The initial condition for the consensus dynamics is then chosen again as in eq.\ref{eq:init_cond_cons}. We run the consensus simulation for 100 bins, we fit the data with the GMM algorithm already described in section \ref{app:GMM}. The separation between the two clusters is quantified through a mass-imbalance observable derived from the GMM model. Let
\begin{equation}
p(x) = \sum_{k=1}^{2} w_k \, \mathcal{N}(x \mid \mu_k, \sigma_k^2)
\end{equation}
be the probability density obtained from the two-component GMM, where $w_k$, $\mu_k$, and $\sigma_k^2$ denote the weights, means, and variances of the components.

We define the negative and positive masses as
\begin{equation}
M_{-} = \int_{-1}^{0} p(x)\, dx,
\qquad
M_{+} = \int_{0}^{1} p(x)\, dx,
\end{equation}
which quantify the total probability weight on the two sides of the origin.

The mass-(im)balance observable is then given by
\begin{equation}
\mathcal{A} = \left| M_{+} - M_{-} \right|.
\end{equation}
\label{eq:mass_balance}

\begin{figure*}[h]
    \centering
    \includegraphics[width=\textwidth]{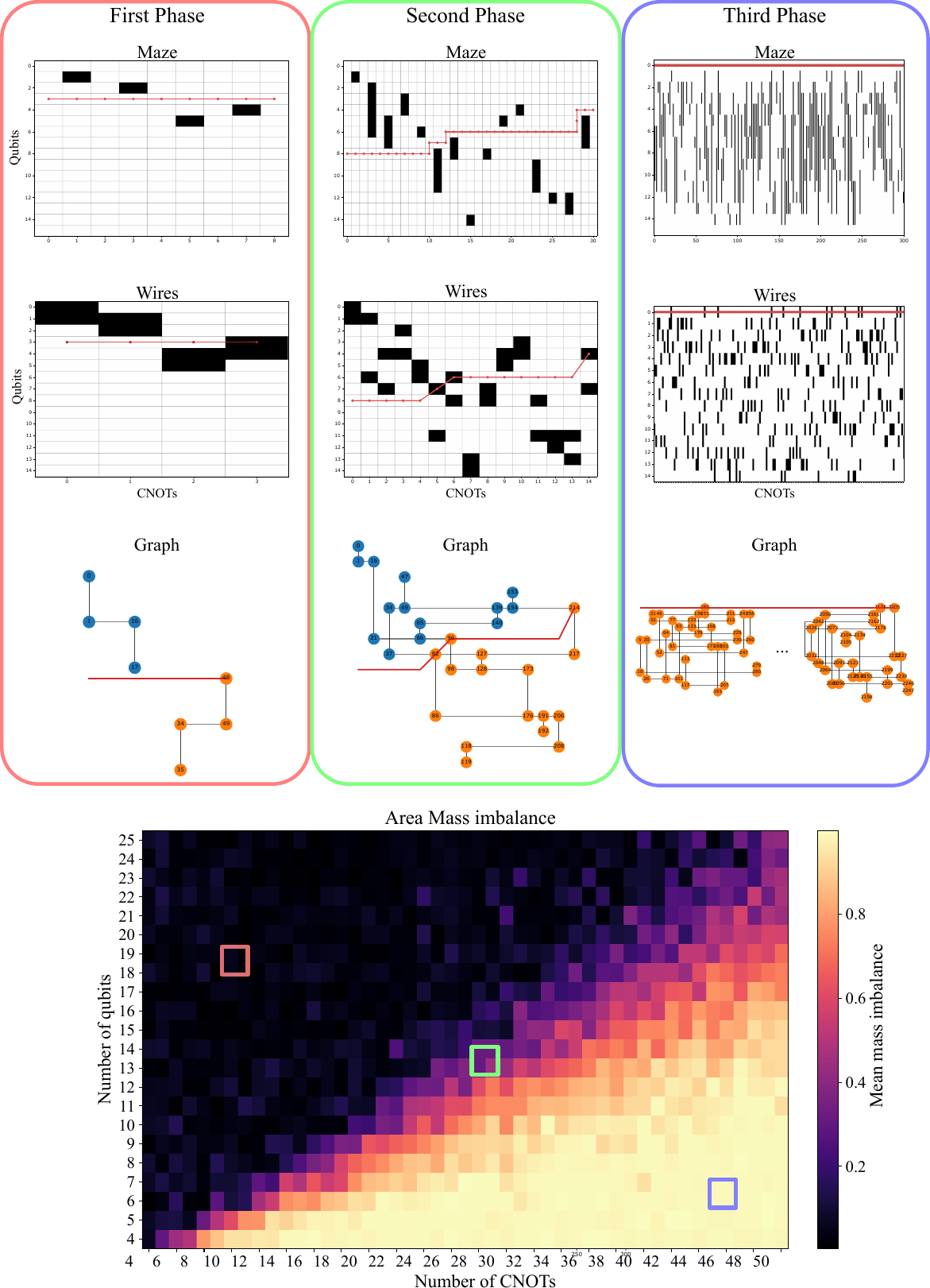}
    \caption{\raggedright Three transport regimes as a function of circuit density. 
    Top: maze, wire, and graph representations for increasing connectivity. 
    In the first phase, the system is split into two disconnected clusters with no transport across the cut. 
    In the second phase, partial connectivity enables transport and mixing between the partitions. 
    In the third phase, the graph becomes fully connected and the partition is lost. 
    Bottom: phase diagram quantified by the mass-imbalance observable $\mathcal{A}$, with $\mathcal{A}\approx 0$ in the separated regime, intermediate values for partial transport, and $\mathcal{A}\to 1$ we have only one partition while the other one is annihilated.
    }
    \label{fig:mass_transport}
\end{figure*}

This quantity provides a simple measure of cluster separation and captures the transport of mass across the interface in a non-monotonic way. In the regime where the two clusters are well separated and remain localized on opposite sides of the interface, the masses balance (eq. \ref{eq:mass_balance}), yielding $\mathcal{A} \approx 0$, which corresponds to the absence of transport. As mixing increases, probability mass is transferred across the interface, leading to an imbalance $M_{+} \neq M_{-}$ and hence $\mathcal{A} > 0$. Finally, in the strongly mixed regime, the two clusters merge into a single effective distribution localized predominantly on one side, so that $M_{+} \to 1$ or $M_{-} \to 1$, and $\mathcal{A} \to 1$. 

The phase transition can therefore be interpreted as a transition between three transport regimes: a separated regime with no transfer, an intermediate regime with partial transport across the cut, and a fully mixed regime in which the distinction between the two partitions is lost.

This is exactly what we observe in Figure \ref{fig:mass_transport}. We can distinguish the three phases in the bottom part of the figure. Plotting examples of circuit of each phase we can see that (i) in the first phase the graph of the two partitions is not connected. (ii) In the second phase we can notice that the graph is split into two partitions and those are connected. (iii) In the third phase the graph is connected however the agent cannot cross the maze in a monotonic way, therefore it creates one single partition.

\section{From network science to quantum computation}

\subsection{Results for simulations of maze cutting at $\gamma=1,3$}
\label{appendix:gamma}

Here we show the simulation results for the metrics $\mathcal{L}_1$, $\mathcal{L}_2$, and $\mathcal{L}_3$, explained in Section \ref{section:phase transition}, but assuming different values of $\gamma$, specifically $\gamma=1$ and $\gamma=3$.
Results are depicted in Figure~\ref{fig:gamma=1,3}.

In the first phase, indicated by the pink square on the left, both best agents are able to traverse the maze horizontally, keeping the number of CNOTs and cells roughly in equilibrium.

\begin{figure*}[t]
    \centering
    \includegraphics[width=0.95\textwidth]{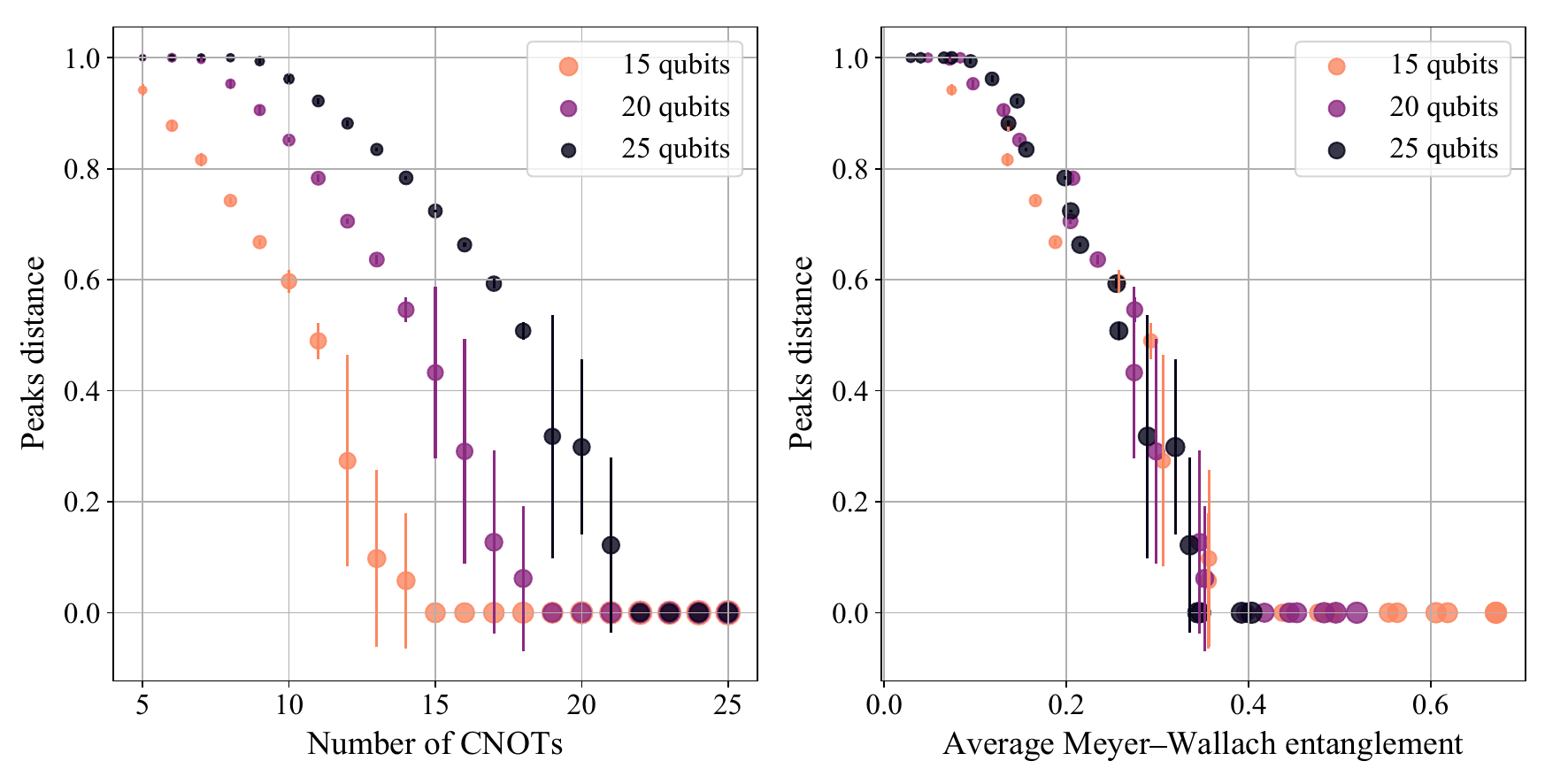}
    \caption{\raggedright The figure shows that the distance between the two peaks of the interpolated distribution is strongly anticorrelated with the average entanglement of the circuit.
    The test was performed on circuits with $15$, $20$, and $25$ qubits and a number of CNOTs ranging from $5$ to $25$.
    In the image on the left, as the number of CNOTs increases, the distance between the peaks progressively decreases until it disappears, indicating the loss of bimodal structure.
    On the right, the data is plotted as a function of average entanglement, and a clear monotonic trend emerges.
    Larger separations between the peaks correspond to weakly entangled states, while highly entangled circuits are associated with peak distances close to zero.}
    \label{fig:meyer-wallach_vs_entanglement}
\end{figure*}

In the second regime, marked by the blue square, the behavior is similar, but in this case both agents cut the maze trivially, along the edges.

In the central region, marked by the green square and indicating the transitional regime, the behavior is completely different.
The agent in (a) favors paths that traverse the maze vertically in the center, in order to minimize $\mathcal{L}_1$ and $\mathcal{L}_2$.
This way, the circuit is cut predominantly vertically rather than horizontally, with the consequence that the phase transition area observed in the image of the metric $\mathcal{L}_3$ increases compared to the case $\alpha=1, \beta=1, \gamma=2$.
In fact, the maze cannot be traversed vertically only if, from the start, the walls cover almost the entire vertical column. 
In this case, the agent is forced to follow a path that points almost immediately to either the top or  bottom edge.
The agent in (b) instead prefers horizontal cuts, but since the maze is not structured to easily allow them in the center, it prefers to cut the maze in a trivial way, namely along the top and bottom edges of the board.

\subsection{Entanglement witnesses}

We have shown that, under suitable conditions, the distribution of the maze walls develops a clear two-peak structure. 
This behavior can be understood in terms of the action of SA algorithm, which rearranges the CNOT gates to form two separate clusters. 
As a result, the corresponding distribution acquires a central lower region and two symmetric maxima located on either side of the midpoint. 
To quantify this bimodal structure, we introduce the distance between the two dominant peaks and relate it to the global entanglement of the circuit, measured by the Meyer-Wallach quantity \cite{brennen2003observable,meyer2001global},
$$
E(\rho)=2\left(1-\frac{1}{N}\sum_k \mathrm{tr}(\rho_k^2)\right),
$$
where $\rho_k$ is the reduced density matrix of the $k$-th qubit and $N$ is the total number of qubits. We should be aware, however, that this only one of the many types of entanglement that we can observe in multi-qubit quantum systems. Moreover, some entanglement witnesses are known to be unreliable in detecting different types of entangled states even for smaller quantum systems, such as two- and three-qubit states \cite{Sauer2021}. We cannot infer from these simulations whether other types of entanglement measures could show different behaviours with respect to the phase transition shown in Figure~\ref{fig:meyer-wallach_vs_entanglement}.

The maximum distance between the peaks is extracted directly from the interpolated distribution after $10^5$ SA steps. 
Local maxima are identified numerically by selecting sampled points whose values are greater than or equal to those of adjacent bins, while boundary points are checked separately to include any edge maxima. 
Denoting $u_{m_1}$ and $u_{m_2}$ as the positions of the two dominant maxima along the rescaled coordinate $u_i$, we define:
$$
d_{\mathrm{peak}}=\left|u_{m_2}-u_{m_1}\right|.
$$
This observable provides a simple geometric measure of the degree of bimodality and allows us to follow the evolution of the separation between the two peaks during the annealing process.

With this definition in hand, we can compare the geometric information contained in the distribution with the entanglement generated in the circuit.
Figure~\ref{fig:meyer-wallach_vs_entanglement} displays that $d_{\mathrm{peak}}$ is strongly anticorrelated with the average Meyer-Wallach entanglement.
The analysis is conducted for circuits with $15$, $20$, and $25$ qubits in a state-vector simulation and for a number of CNOT gates ranging from $5$ to $25$.
We therefore focus on the first-phase region, up to the transient regime, where the relationship between the number of qubits and the number of CNOT operations changes from sublinear to linear.
We can note that, as the number of CNOTs increases, the separation between the two peaks progressively decreases until it disappears, signaling the disappearance of the bimodal structure.
Furthermore, the peak spacing is directly plotted as a function of the average entanglement, revealing clear monotonic behavior.
High values of $d_{\mathrm{peak}}$ are associated with weakly entangled circuits, while highly entangled states correspond to a single-peak distribution.
Interestingly, the data for different system sizes appear to converge on a common curve, suggesting that the peak spacing captures, in a simple geometric way, the accumulation of multipartite entanglement.

\begin{figure*}[t]
    \centering
    \includegraphics[width=0.8\textwidth]{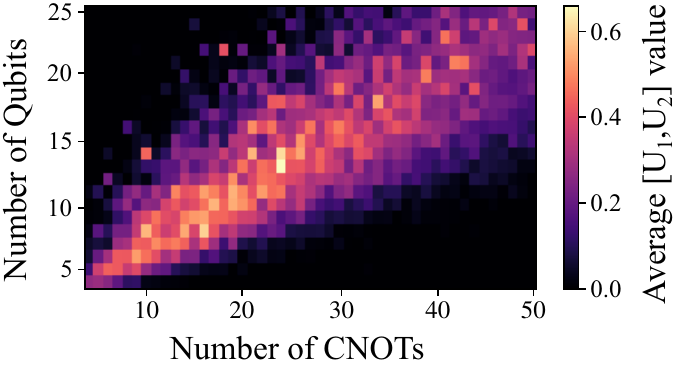}
    \caption{\raggedright The figure shows the average commutator value for $\gamma=2$ as a function of the number of CNOT gates and qubits. 
    The heatmap highlights an intermediate regime of nonzero commutator values, corresponding to the onset of overlap between the two subcircuits, while the commutator vanishes in the regions where the subcircuits remain effectively independent or where the path no longer cuts the circuit.}
    \label{fig:commutator_gamma=2}
\end{figure*}

\subsection{Commutator between partitions}
\label{app:commutator}

We study the commutator between the two partitions of the quantum circuit constructed according to the procedure described above. Denoting \(U_1\) and \(U_2\) as the unitary operators associated with the upper and lower partitions, respectively, we consider:
\begin{align}   
\mathcal{Z} = \sqrt{d} \cdot \mathbb{E}_{U_1, U_2 \sim P(U)}\!\left[\left|\bra{0_N}[U_1,U_2]\ket{0_N}\right|\right]
\end{align}
The prefactor should be considered here as a convenient normalization rather than the consequence of a rigorous derivation. 
Its role is to compensate for the fact that, as the number of qubits increases, the matrix elements between fixed computational basis states tend to become smaller, simply because the dimension of the Hilbert space grows exponentially.

More specifically, for \(N_q\) qubits, the dimension of the Hilbert space is \(d=2^{N_q}\). 
For random Haar unitaries, a matrix element between two fixed basis states is typically of the order \(d^{-1/2}\) \cite{mele2024introduction}. 
Although the set of circuits considered in this work is not directly sampled by the Haar measure, this scaling provides a useful intuition for the dimensional suppression one would expect in a sufficiently random many-qubit unitary. 
From this perspective, multiplication by $2^\frac{N_q}{2}=d^{1/2}$ eliminates the size-dependent leading decay and simplifies the comparison of the commutator amplitudes obtained for different numbers of qubits.

Figure~\ref{fig:commutator_gamma=2} illustrates the results. 
The average commutator over $1000$ simulations is essentially zero in the region where the number of qubits exceeds the number of CNOTs because the two subcircuits do not develop significant overlap. 
In this regime, the subcircuits do not share any qubit wires, so they act independently and their switch cancels out. 
This is consistent with the behavior observed for the third metric $\mathcal{L}_3$ discussed in Section~\ref{section:phase transition}. 

As the ratio between the number of qubits and CNOTs in the circuit decreases, a clear band of non-zero commutator values emerges in the heat map.
This indicates the beginning of overlap between the two subcircuits. 
Indeed, once they share one or more qubit wires, the two parts of the circuit are no longer independent and the commutator acquires non-zero values. 
The structured transition from the dark region to the lighter diagonal band thus marks the regime in which interactions between subcircuits become relevant.

In the second phase, however, the path in the maze representation no longer cuts the circuit into two nontrivial interacting parts. 
Instead, it effectively compares the unitary value of the entire circuit to the identity, so that the commutator trivially reduces to zero. 
This explains why the nonzero regime is confined to an intermediate region, while the commutator reverts to near-zero values outside it.

\end{document}